\documentstyle[12pt,aaspp4]{article}
\def\etal{{\it et al.~}}
\def\ie{{\it i.e.,~}}

\def\eg{{\it e.g.~}}

\begin{document}

\title{Structure and Stability of Keplerian MHD Jets}

%%%%%%%%%%%%%%%%%%%%%%%%%%%%%%%%%%%%%%%%%%%%%%%%%%%%%%%%%%%
\author{Thibaut Lery\altaffilmark{1,2}, Adam Frank \altaffilmark{3}}
\authoraddr{T. Lery\\Dept. of Physics\\ Queen's University,
       Kingston,Ontario, K7L 3N6, Canada}
%\altaffiltext{1}{Depart of Physics, Queen's}
%\altaffiltext{2}{C. E. K. Mees Observatory, and
%Department of Physics and Astronomy,
%University of Rochester, Rochester, NY 14627-0171}
\affil{$^1$ Department of Physics, Queen's University,
Kingston, ON K7L 3N6, Canada}
\affil{$^2$ Dublin Institute of Advanced Studies, 5 Merrion Square,
Dublin 2, Ireland}
\affil{$^3$ Dept. of Physics and Astronomy,
Univ. of Rochester, Rochester, NY 14627-0171}
%%%%%%%%%%%%%%%%%%%%%%%%%%%%%%%%%%%%%%%%%%%%%%%%%%%%%%%%%%%

%\begin{center}{\tt Draft \today}\end{center}

\begin{abstract}
MHD jet equilibria that depend on source properties are obtained using
a simplified model for stationary, axisymmetric and rotating magnetized
outflows (Lery \etal\cite{lery1},\cite{lery2}). The present rotation
laws are more complex than previously considered and include a Keplerian
disc. The ensuing jets have a dense, current-carrying central core
surrounded by an outer collar with a return current. The intermediate
part of the jet is almost current-free and is magnetically dominated.
Most of the momentum is located around the axis in the dense core and
this region is
likely to dominate the dynamics of the jet. We address the linear
stability and the non-linear development of instabilities for our
models using both analytical and 2.5-D numerical simulation's. The
instabilities seen in the simulations develop with a wavelength and growth
time that are well matched by the stability analysis. The modes
explored in this work may provide a natural explanation for knots
observed in astrophysical jets.
\end{abstract}

\keywords{ISM:jets and outflows -- MHD -- Stars: formation -- stability}

%%%%%%%%%%%%%%%%%%%%%%%%%%%%%%%%%%%%%%%%%%%%%%%%%%%%%%%%%%
\section{Introduction}
%%%%%%%%%%%%%%%%%%%%%%%%%%%%%%%%%%%%%%%%%%%%%%%%%%%%%%%%%%

% importance of realistic models for simulations
Observations of collimated outflows from young stellar objects
(YSOs) and active galactic nuclei (AGN) suggest that magnetic
fields may play a central role in the physics of these phenomena.
Despite the large
database of observations, theoretical MHD approaches
have not yet converged on answers to several fundamental questions regarding
the acceleration, collimation and propagation of these jets (See Lery
\etal \cite{lery2} and reference therein). Difficulties investigating the
nature of the jets arise mainly due to the high level of nonlinearity
in the governing MHD equations particularly at critical points.

% The model
The work presented here is based on a simple model for jet launching and 
collimation presented in
previous articles (Lery \etal\cite{lery1},\cite{lery2}, hereafter
Paper I and II).   It yields asymptotic MHD jet equilibria that
account for the properties of the emitting source. The model is
axisymmetric and stationary and in the work presented here
includes more multi-component rotation
laws for the source.  The model assumes the magnetic surfaces possess a
shape which is known a priori inside the fast critical surface,
(we refer to the space within the fast surface as the {\em inner region})
As a first approximation magnetic surfaces were taken to be cones.

The general problem of determining the stationary two-dimensional structure
of magnetohydrodynamic outflows requires the solution of the
equilibrium of forces perpendicular and parallel to the magnetic
surfaces.  One can describe the former by using the transfield or
Grad-Shafranov equation, and the later using the Bernoulli equation
for a polytropic equation of state.
% Equations in the cylindrical region At distances far from
Asymptotically, the jet is assumed to be in pressure equilibrium with
an external medium whose properties are independent of distance. The
transfield and Bernoulli equations are solved both in the inner region
and in the {\em asymptotic
region}, which we also will refer to as the {\em cylindrically
collimated regime}.

% Stability
The stability of the resulting cylindrical equilibria is of major importance,
firstly because instabilities could explain observational features of
jets such as wiggles, knots or helical filaments, and secondly, because
they could be globally disruptive for the outflow. Thus,
in this paper,  we investigate both the
linear stability and the non-linear evolution of our MHD equilibrium
jets. This study will provide a diagnostic for features that should
develop in fully time-dependent simulations of jet propagation.
Such simulations will be the subject of another paper
(\cite{FrankLea1999}).

% GOALS OF THE PAPER AND CONTENTS
Since jet behavior may be directly related to the properties
of the emitting source, the models presented here can provide 
a basis for a better understanding of
jet interactions with an ambient medium as well as propagation and
stability issues in the context of the nature of the source. The main goals of the present
paper are to produce, analyze and study more realistic MHD outflows than simple
``top-hat'' beams (constant density and velocity) used in previous studies
(\cite{Frankea98},\cite{Cenquria99}, \cite{Gardinerea99}, 
\cite{StoneHardee99}).  . 
These new models can then
be used as input for numerical simulations.
How outflows behave on critical surfaces and in the cylindrical regime
constitutes the subject of the first part of the paper.
In \S2, we briefly recall the main features of the model. Jet equilibria
and their properties are presented in \S3. Stability analysis and non
linear developments of the instabilities for the corresponding equilibria
are investigated using ideal MHD computations and numerical
simulations respectively in \S4. Finally, we summarize the results
in \S5.

%%%%%%%%%%%%%%%%%%%%%%%%%%%%%%%%%%%%%%%%%%%%%%%%%%%%%%%%%%
\section{The Model}
%%%%%%%%%%%%%%%%%%%%%%%%%%%%%%%%%%%%%%%%%%%%%%%%%%%%%%%%%%

% the model and its assumptions
The Given Geometry (GG) model is based on the assumption that magnetic
surfaces possess a shape which is known a priori inside the fast critical
surface. For full details on the model and on justifications of the
approximations, we suggest the reader see see \cite{HeyNor89}, and
Paper I and II. Inside the
fast critical point the three critical points are assumed to be aligned
on conical magnetic surfaces. Moreover the angular
distribution of magnetic flux is assumed to be uniform.
These approximations will be relaxed in forthcoming works. 

% Transfield on the Alfv\'en surface
Unlike the Weber-Davis (\cite{wd}) type models, however, the balance of
forces perpendicular to magnetic surfaces is accounted for on the Alfv\'en
surface and at the base of the flow.  On the Alfv\'en surface the force
balance equation becomes what is known as the Alfv\`en regularity
(non-singularity) condition.  Owing to axial symmetry, stationarity, as
well as the flux freezing condition, there exist in the general case
five {\em integrals of motion} that are preserved on any axisymmetric
magnetic surface (denoted $a$).  Two of the integrals, the angular velocity
$\Omega(a)$ and a factor related to the entropy $Q(a)$,  are given as
boundary conditions in the model.  These two input fuctions constitute
assumptions about the nature of the source rotator.

MHD flows have two other critical points which are brought about by the
Bernoulli equation.  There are slow and fast magneto-sonic points both of
which are located where the poloidal velocity equals one of the two magneto-sonic
mode speeds. The fast and slow surfaces, unlike the Alfv\'en surface,
are saddle points, \ie transonic solutions only exist for a certain
relation among the integrals of motion. These relations are obtained
from the criticality conditions that correspond to the vanishing of the 
differential form of the Bernoulli equation at constant $a$ with 
respect to $\rho$ and $r$.

The Alfv\'en regularity condition together with the criticality
conditions determine the three other unknown integrals:
namely the specific energy $E(a)$; the specific
angular momentum $L(a)$; the mass to magnetic flux ratio $\alpha(a)$.
We note that once these integrals are determined
the model gives only an approximate
solution because the transfield equation is not solved everywhere, but
only at a few special places in the {\em inner region}.

% polytropic assumption
We further assume the density $\rho$ to be related to the pressure $p$ by
a polytropic equation of state, $p =  Q(a) \rho^{\gamma}$ where $\gamma$
is the polytropic index. This assumption replaces consideration of energy
balance and is meant to represent simply some more complex heating and
cooling processes (See, for example, Vlahakis \& Tsinganos \cite{VT}
for more general equations of state).

% general definitions
Since the flow eventually becomes fully
collimated far from the source, we use
cylindrical coordinates ($r$,$\phi$,$z$) and assume axisymmetry.  $R$
will be the spherical distance centered on the wind source.
Each flux surface is labelled by the flux function $a(r,z)$ proportional
to the magnetic flux through a circle centered on the axis passing at
point $r$, $z$. The physical flux is $2 \pi a$. We normalize the magnetic
flux to the total magnetic flux $\mathcal A$ enclosed in the jet,
so that $a$ is set
to unity on the outer edge. It is convenient to split the magnetic field
and the velocity into a poloidal part, which is in the meridional ($r$,$z$)
plane, and a toroidal part. The former is denoted by a subscript $P$ while
the latter is just the azimuthal component. We note that the outflow does
not necessarily fill all space from pole to equator.

% Equations
In the {\em inner region},  the system is governed by seven equations:
the Alfv\'en regularity condition; the four criticality conditions
defined at the slow and fast magneto-sonic points; the Bernoulli
equation written at the fast and slow critical points. We note that the
number of equations can be reduced to six, due to the conservation
of the energy on magnetic field lines. Using these equations 
the acceleration of the flow is studied up to the fast critical surface.
When the flow exits the
{\em inner region}, the flow begins to collimate as field lines bend towards the axis.
We do not explicitly follow the collimation process
between the fast critical surface ({\em inner region})
and the {\em cylindrically collimated region}. We note that between these
regions a redistribution of
the energy and angular momentum might occur due to shocks or other
dissipative process. Nevertheless, the integrals of motion are assumed
here to remain exactly the same all along the outflow for simplicity.
Therefore, within the framework of the Given Geometry model,
the asymptotic flow is uniquely determined from first-integrals
obtained from the sub-fast surface regions. In the
{\em cylindrically collimated regime}, the system consists
of the asymptotic forms of the Bernoulli and transfield equations. 
In this region, $z$ goes to $\infty$, the flow density $\rho$ is
smaller than the Alfv\'enic density $\rho_A$, $r$ can be considered 
to be larger than $r_A$, and the gravity becomes negligible.
The full system of equations remains the same as in papers I and II
and is given in Appendix A.

% Input parameters
We note again that in the {\em inner region}, the problem is entirely specified by two
functions of $a$, namely $\Omega(a)$ and $Q(a)$, and also by one constant, the
mass-to-magnetic flux ratio on the axis $\alpha_0$. In the {\em asymptotic
regime}, only one parameter is needed which can be either the external confining
pressure $p_{ext}$, or the axial density $\rho_0$.

% definition of omega
Given the properties of the central emitting object, the model allows one to
compute the dimensionless rotation parameter $\omega\equiv\Omega r_A/v_{PA}$,
where $r_{A}^2 \equiv L/\Omega$ is the Alfv\'en radius and $v_{PA}$ the
Alfv\'enic poloidal speed at the Alfv\'en point
$v_{PA} \equiv \left({{\alpha |\nabla {\bf a}|}/\left(\rho r\right)}
\right)_{A}$. The density at the Alfv\'en point  is given by
$\rho_A \equiv \mu_0\alpha^2$ and the Alfv\`enic Mach number by
$M_A^2 = v_P^2/v_{PA}^2 = {{\rho_A}/{\rho}}$.
More generally the subscript $A$ will refer to values
at the Alfv\'en point. Quantities referring to the two other critical points,
namely slow and fast magneto-sonic critical points, will be indicated
by subscripts s and f respectively. Rotators can be defined as slow,
intermediate or fast according to whether the maximum of $\omega$ is much
less than unity, close to unity, or near its maximum value for of
$(\frac{3}{2})^{3/2}$ (See Paper II for details).

%%%%%%%%%%%%%%%%%%%%%%%%%%%%%%%%%%%%%%%%%%%%%%%%%%%%%%%%%%
\subsection{The Rotation Laws}
%%%%%%%%%%%%%%%%%%%%%%%%%%%%%%%%%%%%%%%%%%%%%%%%%%%%%%%%%%

It is now well known that outflows can be accelerated from an
accretion disk (``Disk wind'', Blandford
\& Payne \cite{bp}, Pelletier \& Pudritz \cite{pellpud}, Fiege \& Henriksen
\cite{fiege}, Contopoulos \& Lovelace \cite{conto}, Ferreira \& Pelletier
\cite{ferr1}, Ferreira \cite{ferr3}, Vlahakis \& Tsinganos \cite{VT}, Lery
 \etal\cite{LHF}), or at the disk-magnetosphere boundary (``X-winds'', Shu
 \etal\cite{shul}, \cite{shu94}).
Therefore, a {\em Keplerian} rotation profile is one of the most realistic description
for rotation of sources producing jets. In paper I and II, we have assumed
a rigid body rotation. Here, we have chosen more realistic rotation laws
by including the {\em Keplerian} rotation in the outermost part of
the outflow.
%%%%%%%%%%%%%%%%%%
\placefigure{fig:OMEGA}
%%%%%%%%%%%%%%%%%%
In Fig.~\ref{fig:OMEGA}, the different profiles of angular velocity used in
our study are
presented. The pure {\em Keplerian} rotation law (dashed line) starts with
a constant rotation close to the axis, as in the rigid body case (dot-dashed
lines), but then follows a {\em Keplerian} profile. The multi-component
(solid line), or {\em multi-fast}, case also starts with a rigid rotation
corresponding, for example, to an axial ordinary wind. The angular velocity
then doubles its value in order to model a source rotating more rapidly than
the star in an intermediate region between
the ordinary wind and the Keplerian disc wind that follows. Note that the
angular velocity is always sub-{\em Keplerian} in the intermediate region.
For all the rotation laws, the axial value of the angular velocity
$\Omega_0$ is set to unity in the figure, and the radius is normalized
to the size of the jet. In order to compare fast and slow rotators,
we will respectively refer to {\em multi-fast} and {\em multi-slow} cases,
the latter one rotating four times more slowly.

%%%%%%%%%%%%%%%%%%%%%%%%%%%%%%%%%%%%%%%%%%%%%%%%%%%%%%%%%%
\section{The Jet Equilibrium}
%%%%%%%%%%%%%%%%%%%%%%%%%%%%%%%%%%%%%%%%%%%%%%%%%%%%%%%%%%

%%%%%%%%%%%%%%%%%%%%%%%%%%%%%%%%%%%%%%%%%%%%%%%%%%%%%%%%%%
\subsection{The Numerical Procedure}
%%%%%%%%%%%%%%%%%%%%%%%%%%%%%%%%%%%%%%%%%%%%%%%%%%%%%%%%%%

In the inner part of the flow, the variables calculated
in the numerical procedure are the radii $r_s$, $r_f$, $r_A$,
and densities $\rho_s$, $\rho_f$, $\rho_A$ at the three critical
surfaces along with the total energy $E$. In its cylindrical regime, the jet
is entirely defined by $r$ and $\rho$.
All the other physical quantities can be derived from
this set. For the numerical calculations, equations have been
reformulated as ordinary differential equations or converted from
algebraic conditions into ODEs as functions of the flux surfaces $a$.

The system consists of eight differential equations, namely the four
equations of regularity on slow and fast surfaces (two on each surface),
the Alfv\'en regularity, the conservation of energy (a total of six for
the {\em inner region}), and the Bernoulli and transfield equations in
their asymptotic form (for the {\em cylindrically collimated regime}).

The numerical solutions are obtained by initiating the integration of
the system from the axis. Given the input parameters $Q_0$,
 $\Omega_0$, $\alpha_0$, $\gamma$, and $\rho_0$, all the critical
positions and densities can be numerically obtained using analytical
formulae (see Paper I). For numerical convenience, we prefer to provide
the axial asymptotic density $\rho_0$ rather than the external pressure.
We further constrain the solution to be super-Alfv\'enic and super-fast-magnetosonic 
on the axis in the asymptotic region. This gives a limiting range of variations for
input parameters and particularly to the axial density $\rho_0$.

%%%%%%%%%%%%%%%%%%%%%%%%%%%%%%%%%%%%%%%%%%%%%%%%%%%%%%%%%%
\subsection{The Input Parameters}
%%%%%%%%%%%%%%%%%%%%%%%%%%%%%%%%%%%%%%%%%%%%%%%%%%%%%%%%%%

The input parameters of the model can be selected so as to qualitatively
reproduce observations. Given the properties of the jet-emitting 
object, \ie its radius $R_{\ast}$, its temperature $T_{\ast}$, 
 the total mass loss rate $\dot M_{\ast}$, the base density $n_{\ast}$,
the magnetic field $B_{\ast}$, the factor $Q_{\ast}$ and $\gamma$,
it is possible to deduce dimensionless parameters
$\overline \Omega$, $\overline Q$, $\overline \alpha_0$.
The parameter  $\overline \alpha_0$ can be a-posteriori
related to the mass loss rate $\dot M_{\ast}$, $R_{\ast}$, and
the magnetic field $B_{\ast}$. Thus we define
$ Q_{\ast}\equiv {2 k T_{\ast} n_{\ast}}/{(m_p n_{\ast})^{\gamma}}$,
$\alpha_{\ast} \equiv {\dot M_{\ast}}/{4 \pi R_{\ast}^2 B_{\ast}}$,
and $\Omega_{\ast}\equiv  \sqrt{{GM_{\ast}}/{R_{\ast}^3}}$.
All quantities are non-dimensionalized to reference values by setting
$\overline Q \equiv Q_{\ast}/Q_{ref}$,
$\overline \alpha_0 \equiv \alpha_{\ast}/\alpha_{ref}$ and
$\overline \Omega \equiv \Omega_{\ast}/\Omega_{ref}$.
The entropy factor $\overline Q(a)$ is assumed to be constant across the jet.
We have also studied models where the jet is hotter along
the symmetry axis than on its outer edges and find that the
results do not significantly
change.

In the present paper, we have chosen to model YSO jets with different
rotation laws using typical values for T-Tauri stars as presented by
Bertout \etal (\cite{bertout})
with $M_{\ast}=0.8 M_{\odot} $ and $R_{\ast}=3 R_{\odot}$. 
At the base of flow, we deduce
the corresponding dimensionless input parameters:
${\overline Q} =0.87$, ${\overline \Omega} =2$,
${\overline \alpha_0} =0.7$, and $\rho_0=5.10^{-7}$.
Major quantities of reference are then given (in CGS)
by $R_{ref} = 10^{15}~cm$, $\rho_{ref}=250~ppc$,
$v_{ref} = 10^{7}~cm~s^{-1}$ for Young Stellar Objects.
The model could be applied to relativistic galactic sources or to quasars,
by adapting the reference values and replacing the external pressure of the
medium by the inertia of an electro-magnetic field outside the light cylinder
in the relativistic regime.

%%%%%%%%%%%%%%%%%%%%%%%%%%%%%%%%%%%%%%%%%%%%%%%%%%%%%%%%%%
\subsection{The Critical Surfaces}
%%%%%%%%%%%%%%%%%%%%%%%%%%%%%%%%%%%%%%%%%%%%%%%%%%%%%%%%%%

% omega
The system is solved on the three critical surfaces in the {\em inner region}.
It has been shown in Paper I that the properties of these surfaces are
related to the rotation parameter $\omega$. As the parameter increases,
the fast and Alfv\'enic surfaces move apart. In the lower right panel of
Fig.~\ref{fig:CRIT}, the rotation parameter is plotted as a function
of magnetic flux. The corresponding input parameters have been given
in the previous section. This figure shows that the rotation
parameter vanishes at the boundaries and reaches
a maximum of 1.6 around $a=0.45$. Thus the jet generated by this flow
will have the properties of a fast magnetic rotator at the location of
the maximum in $\omega$. Another relevant quantity
$\alpha E/\Omega$ is plotted with the rotation parameter. Its minimum
coincides with the maximum in $\omega$. One can
 show that if this quantity possesses a minimum below
its value on the polar axis, the flow will collimate cylindrically
(at least for the region inside the minimum).
This result is in good agreement
with Bogovalov and Tsinganos (\cite{bogovalovT}) who have shown that
there always exist field lines that will collimate cylindrically for
rotating MHD jets. In our case the cylindrical collimation is ensured
due to the pressure of the external medium. 
%%%%%%%%%%%%%%%%%%
\placefigure{fig:CRIT}
%%%%%%%%%%%%%%%%%%

The three critical surfaces and their corresponding densities  are
represented on Fig.~\ref{fig:CRIT}, in the upper and lower left panels
respectively. The surfaces are distorted with the largest deformation
occurring where the rotation parameter reaches a maximum. A relation
between the fast and
Alfv\'enic radii has been found in paper I for the fast rotator case
that states
\begin{equation}
{\frac {r_{A}}{r_{f}}}
\propto
\left[
\beta
\left(3\omega^{\frac{4}{3}}-2\omega^2  \right)^
{\frac{\gamma+1}{2}}
\right]^{\frac{3}{2\left(\gamma-1 \right)}} .
\label{raf2}
\end{equation}
This expression shows that when $\omega$ approaches its maximum
value or when $\beta$ vanishes, the fast point is pushed far from
the Alfv\'en point. It can even be rejected to infinity in the
limiting cases. This is a well known result in the cold plasma limit
(Kennel \etal (\cite{kennel})). In our case, $\beta$ reaches $10^{-2}$ and
$\omega=1.6$ at the point where the critical surfaces are the most elongated.
This behavior has also been obtained in Sakurai (\cite{saku2})
and Belcher \& McGregor (\cite{belcher}).

When rotational effects are small ($\omega\approx0$), \ie close to
the axis, the previous expression becomes
$r_{A}/r_{f} \propto 1- 2 \omega^2 \label{rarf}$
(slow rotator case in paper I). Thus the fast
point gets closer to the Alfv\'enic point in these regions.

% slow surface
The slow surface has the opposite behavior compared to the fast and Alfv\`enic
surfaces. It deflates and gets closer to the source as $\omega$ increases,
as one would expect from $r_s/r_A \propto (r_A/r_f)^{1/3}$ (Paper I). In
fact in the fast rotator regime, the slow mode speed gets closer to the
sound speed, and the slow mode acquires the character of a sound wave
guided along the field line.
%conclusions
Thus {\em the fast magneto-sonic and Alfv\`en surfaces strongly inflate
when rotation increases or the flow becoming cold. On the
other hand, the slow surface gets smaller with increasing $\omega$}.

%densities
The lower left part of the figure clearly shows the trend
of decreasing densities from the axis to the maximum in $\omega$.
At the fast point for fast rotators (see paper I),
\begin{equation}
{\frac {\rho_{f}}{\rho_{A}}} \propto
\omega^{-2/3}\left(\frac {r_{A}}{r_{f}}\right)^2 .
\label{rhofa2}\end{equation}
The square of the last term explains the decrease of
four decades in density.
%first integrals
As stated earlier,  all the other variables of the problem, such as
the first integrals of the motion $E$, $L$ and $\alpha$ can be deduced
from the positions and densities of the critical points. For example,
Fig.~\ref{fig:CRIT} shows that the total angular momentum
reaches a maximum  inside the jet and not on the outer edge.
Thus {\em most of the dynamics will be internal} since this quantity
is conserved along the flow.

%%%%%%%%%%%%%%%%%%%%%%%%%%%%%%%%%%%%%%%%%%%%%%%%%%%%%%%%%%
\subsection{The Asymptotic Equilibrium}
%%%%%%%%%%%%%%%%%%%%%%%%%%%%%%%%%%%%%%%%%%%%%%%%%%%%%%%%%%

\subsubsection{The Numerical Solutions}

Here we compare the influence of the rotation laws on the 
cross-sectional distributions in MHD
jets.  The quantities that define the jet in the cylindrically collimated regime
are plotted in Fig.~\ref{fig:ASYMP} for pure {\em Keplerian},
{\em multi-fast} and constant rotations.
The $z$ and $\phi$ components of velocity and magnetic
field are represented together with the density $\rho$ and the net
electric current $I_C$, as functions of the relative radius (normalized to
the jet radius). The length scale is the jet radius. The density is
normalized to its value on the jet axis $\rho_0$.  The non-dimensional
velocities refer to the fast magnetosonic velocity
$v_f^2 = c_s^2 + v_A^2$ on the axis, $c_s$ being the sound speed.
The magnetic field is normalized to $\sqrt{\rho_0}\ v_f$.

%%%%%%%%%%%%%%%%%%
\placefigure{fig:ASYMP}
%%%%%%%%%%%%%%%%%%

%Velocity
The poloidal velocity $v_z$ increases from the axis
to the outer edge for the rigid rotation case, and slightly decreases
for the pure {\em Keplerian} case. For the {\em multi-fast} case, however,
$v_z$ peaks where the rotation parameter is maximum, approximately in the middle
of the jet. Therefore in the latter case,
the fastest part of the jet is neither on the axis or on the outer edge,
but inside the jet itself. The azimuthal velocity $v_\phi$ follows the
same trend but with a magnitude several orders smaller than the poloidal
component. If $v_z$ is of the order of several hundreds
of $km~s^{-1}$, $v_\phi$ is only about a few $km~s^{-1}$.

%Magnetic field
The azimuthal component of the
magnetic field always dominates the poloidal part except at the axis. Thus for these
parameters the jet's field is highly twisted. The azimuthal component
$B_\phi$ follows approximatively a ${1}/{r}$ law in the outer part.

%density
As one moves towards the axis from the edge of the
jet magnetic pinch forces or {\em hoop stress} become increasingly important.
In order to maintain an equilibrium the
gas pressure must balance the hoop stresses. Thus we see large pressure
and density gradients in near the axis. We denote the high density
region centered on the axis as the {\em core}, and the lower density outer
regions as the  {\em collar}. Note that the bulk of the jet's momentum
resides in the core. Hence this portion of the beam will penetrate more
easily into the ambient medium during the jet's propagation while the
collar will be more strongly decelerated. Thus we expect that even if
the relative velocity is smaller close to the axis, the central part
of the flow will propagate faster.

%Current
The last quantity represented in Fig.~\ref{fig:ASYMP} is
the "poloidal electric current" through a circle centered on the axis
and extending out to a magnetic surface $a$.
This quantity is given by $I_C(a)=-r B_{\phi}/\mu_0$ (The physical poloidal
current is $I_{phys}=-2\pi I_{C}$).
While the current is always increasing for a constant rotation jet,
in the {\em Keplerian} and multi-component models it reaches
a maximum in the middle of the beam
and vanishes at the
outer boundary. Therefore a return current flows back
inside the jet in these cases. This will be studied in more
detail for the {\em multi-fast} case in the next section.

%ratio
The kinetic to Poynting flux ratio (not shown here) can be
calculated as a function of the relative radius. We find that
only the central part of the asymptotic outflow has a kinetic
energy flux strongly dominating the Poynting flux. Thus
away from the axis a non-negligible part of the magnetic energy is not
transformed into kinetic energy.

Thus {\em rigid body rotation jets are characterized by a dense,
current-carrying core having most of the momentum, surrounded by
a tenuous current-free envelope, dominated by the azimuthal magnetic
field. Keplerian and multi-fast cases also exhibit a central
current-carrying core but are surrounded by a denser collar
and carry an internal return current}.

\subsubsection{An Approximate Analytical Solution}

By simplifying the asymptotic equations, it is possible to
obtain approximate analytical solutions in the region where
the jet is cylindrical. Away from the axis, where the gas pressure
is negligible with respect to magnetic pressure, the transfield
equation can be integrated to yield
$\Omega^2 r^4 \rho^2 \mu_0^{-1} \alpha^{-2}= constant$.
Then the density can be expressed as a function of $r$ and of
the first integrals
\begin{equation}
\rho(r) \approx {C \alpha}/{\Omega r^2} ,
\label{rhoTHEO}
\end{equation}
where $C$ is a constant. In the present work, the angular velocity
$\Omega(r)$ is given as an initial condition, while the mass to magnetic
flux ratio $\alpha(r)$ is part of the solution.
If $\alpha(r)$ and $\Omega(r)$ are constant,
the density drops as $r^{-2}$. This is the behavior that we get for
the rigid body rotator away from the axis. On the other hand, for increasing
$\alpha$, or decreasing values of $\Omega$, the density can increase
as obtained for the {\em multi-fast} rotator in the outer regions of the jet.
Close to the axis, \ie $r\ll 1$, the gas pressure dominates and the
density approaches a constant, the transfield equation reducing
approximately to $Q \rho^\gamma=constant$.

The above result allows one to derive similar approximate formulae for the
toroidal components of the magnetic field and velocity which respectively
reduce to  $\vert B_\phi \vert\approx \Omega \rho r / \alpha$, and
$\vert v_\phi \vert \approx \rho\Omega r/\mu_0\alpha^2$.
Combined with Eq.~\ref{rhoTHEO}, the equations become
\begin{equation}
\vert B_\phi(r) \vert \approx {C}/{r} \quad ,
\quad \vert v_\phi(r)\vert \approx {C}/{\alpha} .
\label{BphiTHEO}
\end{equation}
It follows that the net asymptotic electric current is approximately
given by $I=2\pi r \vert B_\phi\vert/\mu_0\approx constant$. Therefore,
in the region where magnetic pressure dominates over gas pressure, the
asymptotic current will be constant.
 Finally, one can deduce the asymptotic poloidal velocity of the flow
$v_z(r)=(\alpha/\rho r)da/dr$, that becomes
\begin{equation}
v_z(r) \approx \Omega r / C .
\label{VzTHEO}
\end{equation}

%%%%%%%%%%%%%%%%%%
\placefigure{fig:THEO}
%%%%%%%%%%%%%%%%%%
Fig.~\ref{fig:THEO} shows the approximate analytical solutions
together with the solution for the {\em multi-fast} case. The density
and the poloidal velocity are well reproduced in the intermediate 
region where the magnetic pressure dominates. 
The toroidal component of the magnetic field
is well described by the formula only far from the axis.
Finally, the toroidal component of the velocity is similar in the
analytical and the numerical solutions.

In theory, by using this approximate solution, it may be possible to
deduce the properties of an outflow and, therefore, of the
emitting object directly from quantities observed in the jets.
A rough estimate of $\alpha$ could be deduced from the toroidal
component of the velocity. Combined with an accurate measure of
the density and of the asymptotic axial velocity, it would give
the angular velocity. Using the model, the five first integrals 
could be estimated and would allow a characterization of the 
source itself. Moreover, just from the
profile of the density, it would be possible to deduce the mass to
magnetic flux ratio, $\alpha(r)$, for a given rotation law. Thus
{\em a precise measure of the mass density in jets may be a key
point in order to deduce the properties of the source at its base}.

%%%%%%%%%%%%%%%%%%%%%%%%%%%%%%%%%%%%%%%%%%%%%%%%%%%%%%%%%%
\subsection{The Current along the Jet}
%%%%%%%%%%%%%%%%%%%%%%%%%%%%%%%%%%%%%%%%%%%%%%%%%%%%%%%%%%

%%%%%%%%%%%%%%%%%%
\placefigure{fig:CURRENT}
%%%%%%%%%%%%%%%%%%

In Fig.~\ref{fig:CURRENT}, we present the net electric current
on the Alfv\'enic and fast surfaces and in the {\em cylindrically
collimated region}, in order to have a global picture of the current
circulation in the jet. Only the {\em multi-fast} case is presented
here since the {\em Keplerian} case presents similar trends.
The current first increases outwards from the axis,
then reaches a plateau where the magnetic pressure dominates, as
shown in the previous section. Finally, it decreases in the outer
part of the jet, and almost vanishes on the outer boundary.
The current density is negative in the outer part of the jet and
therefore the direction of the current is opposite to the one in the
axial region. Thus {\em there exists a strong current in the core and a
return current in the collar, the intermediate part of the jet being
almost current-free}.

We note that the
current is never strictly zero on the outer boundary, and therefore a small
part of the current may flow back either on the external surface of the
jet, or in the external surrounding medium. This result is similar to the
force-free field model of Lynden-Bell (\cite{lynden}), where the current
circulates inside the jet itself, and where the magnetic field lines are
anchored in a differentially rotating accretion disc.

%%%%%%%%%%%%%%%%%%%%%%%%%%%%%%%%%%%%%%%%%%%%%%%%%%%%%%%%%%
\subsection{Comparisons with Previous Works}
%%%%%%%%%%%%%%%%%%%%%%%%%%%%%%%%%%%%%%%%%%%%%%%%%%%%%%%%%%

The characteristics of magnetized outflows at large distances
from the central object have been addressed by a variety of studies.
Many results of the present study are in good agreement with
these works. For example, as found by Beskin \etal (\cite{beskin}),
a jet with a zero total electric current has its angular
velocity vanishing at the jet boundary, and its axial regions dominated
by kinetic energy. Concerning the current, Bogovalov (\cite{bogovalov}) has
shown analytically that there always exists a field line in the outflow
which encloses a finite total current, and consequently, the
asymptotics always contain a cylindrically collimated core. This is
precisely what we find in the axial region of our jets, regardless of the
presence of any external medium that would ensure cylindrical collimation
anyway. As with Ferreira (\cite{ferr3}), it is possible to show, as it is
done in paper I, that the minimum mass loss rate has a lower limit and can
not be arbitrarily small. We also agree with Ostriker (\cite{ostriker})
and Lery \etal(\cite{LHF}) who conclude that the optical jet may 
represent only the densest part of the total outflow. 

%Low Mach number
We obtain fast magnetosonic Mach numbers, (which also corresponds to the
Alfv\'enic Mach number on the axis), between 2 and 4. This range
corresponds to what Camenzind (\cite{cam182}) has found  for his model
for low-mass protostellar objects. The corresponding jets have low fast
magnetosonic Mach-numbers $M_A\simeq 2$. By taking into account an
accretion disc around the stellar magnetosphere, Fendt \&
Camenzind (\cite{fendt}) also find a fast magnetosonic Mach-number
$\approx ~2.5$. However, this does not seem to be a general statement
about MHD jets since there exist models with larger values
(Sauty \etal(\cite{sautytsing}), Trussoni \etal(\cite{trussoni})).

%Comparison with Shu
Shu \etal(\cite{shu95}) have studied magneto-centrifugally driven
flows from young stars including their structure at large distances
from the source when they collimate into jets. When $R \gg 1$ and
$r$ is at least moderately large also, they have found that the density
distribution can be given by  $\rho \rightarrow K(r)/\alpha(r) r^2$
(their Eq.4a) where $K(r)$ is an arbitrary (but slowly varying) positive
function of $r$. This result, and the variations of their asymptotic
velocity, are in good agreement with Eq.~\ref{rhoTHEO},
with $K(r)$ being $C/\Omega(r)$ in the present model. Despite the
similarity of the analytical results, the functions in Shu \etal
do not correspond to the {\em multi-fast} (nor the {\em Keplerian}) case.
Note that our model does not describe the physical processes occurring at
the source itself, \ie at the surface of the disk or the disk-star boundary.
Thus one may be able to link our results to models such as Shu's for the 
generation of the outflow at its source.

%%%%%%%%%%%%%%%%%%%%%%%%%%%%%%%%%%%%%%%%%%%%%%%%%%%%%%%%%%
\subsection{The Astrophysical Consequences}
%%%%%%%%%%%%%%%%%%%%%%%%%%%%%%%%%%%%%%%%%%%%%%%%%%%%%%%%%%

The model makes it possible to obtain MHD jet equilibria with
multiple components, \ie a central dense core surrounded by a
collar. Characteristic physical quantities are close to
those obtained observationally, with, for example, velocities
of several $10^2~km~s^{-1}$ for YSOs. We note that there exists
a broad range of solutions to the equations that possess favorable
characteristics for compararisons with observations.
Moreover, the properties of the different parts of the jet
can be directly related to the properties of the source.
Therefore the ensuing equilibria can be used to simulate
numerically the propagation of the jet into an external medium,
more realistically than with a ``top-hat'' distributions for velocity
and density. By comparing astrophysical observations with
the results of different simulations of jet propagation
could ultimately provide a tool for deriving source
properties.

%%%%%%%%%%%%%%%%%%
\placefigure{fig:BETA}
%%%%%%%%%%%%%%%%%%
Another important issue that must be addressed is the effect
of ambipolar diffusion in our model, and the estimation of the
distance where it should become effective. In a recent work by
Frank \etal (\cite{frankambi}), it has been shown that
the ambipolar diffusion time-scale  was related to
the plasma $\beta$ parameter, $\beta \equiv P_g/P_B$, by
\begin{eqnarray}
t_{ad} & = & 28,904
\left( { n_n \over 10^3 cm^{-3}} \right)
\left( { r_{jet} \over 10^{15} cm} \right)^2
\left( { 10^4 K \over T_{jet}} \right)
\left( {\beta \over \beta + 1} \right)  ~ y .
\label{tad}
\end{eqnarray}
In this equation, the ambipolar diffusion time-scale is given in years,
$n_n$ is the number density of neutral particles,
and $T_{jet}$ is the temperature.
When $\beta \rightarrow \infty$, the term $\beta/(\beta + 1)$
tends to unity, while when $\beta \ll 1$, the ambipolar diffusion
time-scale can be effectively reduced. In the present model,
the plasma $\beta$ parameter is given  by
\begin{equation}
\beta(a) = {{2 \gamma}\over{ \left (\gamma-1\right )}}
 {{Q \rho_A^{\gamma-1}} \over {v_{PA}}^2} .
\label{beta}
\end{equation}
As shown in Fig.~\ref{fig:BETA}, $\beta$ can become quite low in some
part of the jet. The most dramatic case is the {\em multi-fast} case,
where value around $10^{-2}$ can be reached in the intermediate part
of the jet,
the core and the collar having a plasma parameter larger than unity.
Therefore, the jet is magnetically dominated only in the intermediate
zone, precisely where ambipolar diffusion can be effective. For all
the other cases, the average value of the parameter is about unity.
From observations, typical values for YSO jets are:
$10^3 \le n_n/~cm^{-3} \le10^4$;
$1 \times 10^{15}~ \le r_{jet}/~cm \le 5 \times 10^{15}$;
$5 \times 10^3 ~K < T < 3 \times 10^4 ~K$
(Baccioti \& Eisloeffel \cite{BacEis99}).
For jet parameters in the middle of the expected range of
variation, we find $t_{ad}$ of order $10^5$ to $10^4$ y. Thus the
dynamical time-scale for YSO jets is $t_{dyn}=10^4~-~10^5$ y
(Reipurth \etal\cite{Reipurth1997}, Eisloeffel \& Mundt \cite{EisMun1997}).
This is of order of, or greater than, the ambipolar diffusion time:
$t_{dyn} > t_{ad}$. If we consider length scales,
the distance $D_{ad}$, where ambipolar diffusion becomes effective
is approximatively between $0.1 ~pc$ and $3 ~pc$,
for jet velocities around $10^2~km~s^{-1}$. This is the
range of distances where the magnetic field might be altered
or at least where internal configurations should evolve.
Such an effect should reduce the internal {\em hoop stress} 
that would normally
help the jet to remain collimated. It could also reduce instabilities
as it will be shown in the next section. Hence, parsec scales YSO jets
should naturally be less collimated when $t_{dyn}\approx10^4~-~10^5$ y.

Finally, the present jet equilibria show large gradients
in velocity, magnetic field and density. Such configurations
should give rise to instabilities that could change internal
structures of jets or even disrupt them. This is the subject
of the next section.

%%%%%%%%%%%%%%%%%%%%%%%%%%%%%%%%%%%%%%%%%%%%%%%%%%%%%%%%%%
\section{The Stability Analysis}
%%%%%%%%%%%%%%%%%%%%%%%%%%%%%%%%%%%%%%%%%%%%%%%%%%%%%%%%%%

Current-carrying jets with a helical fields 
are susceptible to pressure-driven (PD), Kelvin-Helmholtz (KH) and magnetic
instabilities driven by the electrical current (Current Driven, CD). The latter
modes are due to field aligned electrical currents.
Generally instabilities will be a mixture of CD, PD and KH modes
with the contributions from each type of mode being difficult to estimate.
Stability properties of such MHD jets have been
investigated only recently (Appl \& Camenzind \cite{applcam},
Appl \cite{appl}, Lery \cite{lery3}, Appl \etal\cite{lba}).
These instabilities could play an important role in various observed
morphological structures such as wiggles, (\eg for quasars:
Krichbaum \etal\cite{krichbaum+}, Feretti \etal\cite{feretti},
for YSO jets: Schwartz \& Greene \cite{schwartz}), knots
(\eg for YSO jets: Ray \etal\cite{rayetal}, Raga \& Noriega-Crespo
\cite{raga}, Rosado \etal\cite{RRA}), and filaments (\eg for AGNs:
Biretta \cite{biretta}, Bahcall \etal\cite{bahcall},
for YSO jets: Elmegreen \cite{elmegreen}, Dutrey \etal\cite{dutrey}).

%%%%%%%%%%%%%%%%%%%%%%%%%%%%%%%%%%%%%%%%%%%%%%%%%%%%%%%%%%
\subsection{The Linear Stability Analysis}
%%%%%%%%%%%%%%%%%%%%%%%%%%%%%%%%%%%%%%%%%%%%%%%%%%%%%%%%%%

In order to study linear development of instabilities,
we adopt the standard (temporal) approach
where the axial wavenumber $k$ is real and the
imaginary part of the complex frequency corresponds to growth rate
$\Gamma$, growth time being $\tau=\Gamma^{-1}$. Wavenumbers are
given in units of inverse jet radius and growth rate is normalized
to the inverse Alfv\'{e}n time. A synopsis of
the method is presented in Appendix B.

%%%%%%%%%%%%%%%%%%%%%%%%%%%%%%%%%%%%%%%%%%%%%%%%%%%%%%%%%%
\subsubsection{The Pinch Mode $m=0$ \label{sec:m=0}}
%%%%%%%%%%%%%%%%%%%%%%%%%%%%%%%%%%%%%%%%%%%%%%%%%%%%%%%%%%

Stability analysis of magnetic configurations derived from the Given Geometry
model have been studied in Lery (\cite{lery3}) for rigid rotators. 
We report here the main results. It has
been shown, in such a case, that the pinch, or sausage, mode ($m=0$) dominates
over other $m$-th order modes.  This is mainly due to the
large gradients of the density and of the magnetic field that 
create strong PD and CD instabilities respectively.

The results of (\cite{lery3}) show that pure CD modes 
can develop on rapid, \ie Alfv\'{e}n, time scales. Moreover fast rigid
rotators are more unstable than slow ones.
%%%%%%%%%%%%%%%%%%
\placefigure{fig:m0}
%%%%%%%%%%%%%%%%%%
In Fig.~\ref{fig:m0}, we have plotted the dispersion relation for these
two cases for the pinch mode $m=0$. By studying a large range of angular
velocities, Lery (\cite{lery3}) demonstrated that faster
rotation rates produce larger growth rates and smaller wavenumber
cut-offs. This is clearly seen in the figure, where the fast rotator is
the most unstable for large wavelengths (\ie small wavenumbers), but also
presents a smaller cut-off in wavenumbers. The maximum growth rates and
their corresponding wavenumbers are reported in table~\ref{table:m0}.
%%%%%%%%%%%%%%%%%%
\placetable{table:m0}
%%%%%%%%%%%%%%%%%%
The unstable modes that should grow the most rapidly have a wavelength about
3 jet radii for the fast rotators and half the jet radius for the slow
rotators. These results can be applied to observations of real YSO jets.
For HH34 the largest wavelength above corresponds approximately to the minimum
knot separation of $3.4~r_{jet}$ as given by Burke \etal ({\cite{burke}).
In the case of HH111 however, the separation is larger,
about $11~r_{jet}$ (Reipurth \cite{reipurth}, Morse \etal\cite{morse}).
Thus model seems not to fit to the HH111 case.  If the jet diameter
is not defined as the full width of the jet, $2~r_{jet}$ as we have done,
but as the observationally deduced knot width then the application appears
better. This would correspond, in our model, to the denser, axial core.
In such a case, the jet diameter would be smaller by a factor of 3 or 4
and knot separation would be about $9$ to $12~r_{jet}$, thus accounting
for the larger separation of knots of HH111.

We note also that the number of knots for HH111 is about 13 (Reipurth
\cite{reipurth}) corresponding to an approximate length of the knotted section
of the jet of $4\times10^{17}~cm=0.13~pc$, (given a knot separation of
$3~r_{jet}=3\times10^{16}~cm$.  This is the distance where
{\em ambipolar diffusion} could become effective in the jet, as pointed out
earlier in the present paper. Therefore, at $0.1~pc$ from the source,
the internal configurations should begin to evolve leading to decreases in pressure and
magnetic field gradients. Consequently pressure driven and current
driven instabilities should be less important. This might be the reason
for the disappearance of the knots at this distance from the source.

Thus, {\em regardless the type of rotators, the domination of
axisymmetric modes and their corresponding wavelengths suggest that
these instabilities could be at the origin of the knotted structures
seen in a large
number of jets as seen, for example, in HL Tau, HH1, HH30 and HH34
(Ray \etal\cite{rayetal}), or in HH11 and HH311 (Rosado \etal\cite{RRA})}.

This potential origin for knots has already been proposed by several authors
(Burke \etal\cite{burke}, Bodo \etal\cite{bodo}, Micono \etal\cite{micono}).
These studies focused on Kelvin-Helmholtz instabilities.
Todo \etal (\cite{todo}), however, have argued that the pure
KH instabilities could not alone produce the knots. We propose that
{\em the knots may be due to the combined effects of the KH, CD, and PD
instabilities, with a domination of the PD and MHD instabilities}.

%%%%%%%%%%%%%%%%%%%%%%%%%%%%%%%%%%%%%%%%%%%%%%%%%%%%%%%%%%
\subsubsection{The Helical Mode $\vert m \vert=1$}
%%%%%%%%%%%%%%%%%%%%%%%%%%%%%%%%%%%%%%%%%%%%%%%%%%%%%%%%%%

As stated above an MHD outflow can exhibit Kelvin-Helmholtz, current-driven, and
pressure-driven instabilities. In general, it is not possible to
entirely separate these modess. Only with very simple and idealized
configurations one can hope to understand the physics
driving individual modes.
In this section we restrict ourselves to current-driven modes,
in order to study their effects on our models. We
use the results obtained by Appl \etal (\cite{lba}). They
have approximated the MHD jet by an infinitely long cylindrical
outflow of a perfectly conducting plasma. The jet is supposed to have
constant density and velocity, as well as negligible thermal pressure
and rotation. Consequently, Kelvin-Helmholtz instabilities arise only
due to the vortex sheet at the jet boundary. Moreover, pressure-driven
instabilities are excluded by considering cold jets.
The simplifications differ from our present model, \ie we have velocity
and pressure gradients. Nevertheless it allows
to get a {\em rough estimate} of the maximum growth rate and wavenumber
for each value of the axial pitch. Eventually the ensuing characteristic
wavelengths can be calculated.

Appl \etal (\cite{lba}) have shown that CD helical modes generally dominate
in thier jet configurations.
They found that the magnetic pitch, $P = rB_z/(r_{jet}B_\phi)$,
and in particular its value on the axis, $P_0$, essentially determines
the growth rate of the helical instabilities in the case of small values
of the pitch.
%%%%%%%%%%%%%%%%%%
\placefigure{fig:muGK}
%%%%%%%%%%%%%%%%%%
We have plotted the pitch in Fig.~\ref{fig:muGK} (left panel) as a
function of the relative radius, for {\em Keplerian}, {\em multi-slow} and
{\em multi-fast} rotators. The axial values of the pitch
functions were deduced from the models presented above.
Following the same procedure as
Appl \etal (\cite{lba}), it is possible to derive the
maximum growth rates and wavenumbers as functions of the axial pitch for our jets
(right panel in Fig.~\ref{fig:muGK}).
Results are reported in table~\ref{table:m-1}.
%%%%%%%%%%%%%%%%%%
\placetable{table:m-1}
%%%%%%%%%%%%%%%%%%
The most unstable magnetic configurations correspond to the {\em multi-slow}
case which has a very small wavelength. We note that the
{\em Keplerian} and {\em multi-fast}
cases have maximum growth rates that are still quite large.

One should note that the maximum growth rates given here have been
obtained for a particular value of medium external pressure.
An increase of the external
pressure increases the central pitch $P_0$ which reduces
$\Gamma_{max}$ by a factor of 10. For large external pressure,
the pinch mode can be of the same order or larger than the helical
mode. A 3D simulation would be needed in order to study
the development and the evolution of these instabilities in detail.

%%%%%%%%%%%%%%%%%%%%%%%%%%%%%%%%%%%%%%%%%%%%%%%%%%%%%%%%%%
\subsection{The Non-linear Evolution}
%%%%%%%%%%%%%%%%%%%%%%%%%%%%%%%%%%%%%%%%%%%%%%%%%%%%%%%%%%

%%%%%%%%%%%%%%%%%%%%%%%%%%%%%%%%%%%%%%%%%%%%%%%%%%%%%%%%%%
\subsubsection{The Kink Modes}
%%%%%%%%%%%%%%%%%%%%%%%%%%%%%%%%%%%%%%%%%%%%%%%%%%%%%%%%%%

Non-linear computations of the development of the CD modes have
been carried out by Lery, Baty \& Appl (\cite{lb}) using a cylindrical
evolution code. These calculations show that like the linear evolution,
non-linear instabilities
also develop on rapid time scales and therefore should affect
internal jet structures. Moreover the non-linear
behavior of the jet equilibrium is highly sensitive to the structure
of the initial magnetic configuration. In the present case where
the pitch function is increasing, the dominant mode ($m=1$) is
most probably resonant at half the relative radius of the jet.
A current sheet should form at this locus producing a kinked inner cylinder
due to helical distortion. Reconnection and turbulence
in such a configuration should occur and accelerate particles on
this surface. The ensuing saturated configurations may look like
a hollow cylinder. For AGNs, such a structure may be related to the formation of
the VLBI knots, which apparently are non-axisymmetric features
within the flow channel (Krichbaum \etal\cite{krichbaum+}).
Finally, a high current density forms along the jet axis 
in these calculations.  This should give rise to
radially localized dissipation which could be a potential heating 
mechanism within the jet core.

%%%%%%%%%%%%%%%%%%%%%%%%%%%%%%%%%%%%%%%%%%%%%%%%%%%%%%%%%%
\subsubsection{The Simulations}
%%%%%%%%%%%%%%%%%%%%%%%%%%%%%%%%%%%%%%%%%%%%%%%%%%%%%%%%%%

%%%%%%%%%%%%%%%%%%%%%%%%%%%%%%%%%%%%%%%%%%%%%%%%%%%%%%%%%%
\paragraph{The Numerical Method}
%%%%%%%%%%%%%%%%%%%%%%%%%%%%%%%%%%%%%%%%%%%%%%%%%%%%%%%%%%

A multi-dimensional ($2.5$-D) simulation of a {\em multi-fast} MHD jet
have been performed using a MHD TVD code in cylindrical symmetry
(see Ryu, Jones \& Frank \cite{RJF} and Ryu \etal\cite{Ryu1995-2Dcyl}).
The simulations are initiated with the
cylindrically collimated jet equilibrium traversing the length
of the computational domain ($256~\times~1024$
zones), the jet radius being 64 zones. The equilibrium is then
continuously injected at $z=0$ boundary of the grid. The radius of the
jet was chosen to correspond to a YSO jet
of $10^{16}~cm$. The maximum density of the initial jet was $250~ppc$,
and the central velocity was $10^{7}~cm~s^{-1}$. The pressure
outside the jet was imposed by the model through the pressure balance
at the outer boundary. The jet is surrounded, in
the present simulation, by a magnetized medium that has a small poloidal
field and no toroidal component.

As seen previously, the present magnetic configuration is always
unstable to both the pinch and the helical (or kink) modes. Since our
simulation is axisymmetric, it is only possible to track the pinch modes.
In order to study the kink modes 3D simulations are required.
This will be the subject of a forthcoming paper.

%%%%%%%%%%%%%%%%%%%%%%%%%%%%%%%%%%%%%%%%%%%%%%%%%%%%%%%%%%
\paragraph{The Results}
%%%%%%%%%%%%%%%%%%%%%%%%%%%%%%%%%%%%%%%%%%%%%%%%%%%%%%%%%%

We note first that the equilibrium remains stable over a long period.
No preferential perturbation has been used
in order to destabilize the simulation. As the jet propagates numerical
noise generates the instabilities. Two types of regularly spaced features
arise. One has a small wavelength and is located close to the axis. The
second presents a larger wavelength. the second mode is wider and situated
at the interface between the jet and the ambient medium
as illustrated in fig.~\ref{fig:simDRAWING} and fig.~\ref{fig:simB}
where the density, and the poloidal and toroidal components of the
magnetic field are shown respectively.

In a companion paper (Frank {\it et al} 1999) we have performed simulations
which follow the propagation of the jet models presented here.
These simulations also show instabilities developing in the beam which
have identical wavelengths. The instabilities are initiated
where the beam interacts strongly with ambient medium via the propagation.
Here the instabilities take longer to develop as they are generated via low
level noise.  Note the form of the instability at the axis.  Detailed
inspection of the simulations show a strong axial flow (a pinch).
When material reflects off the axis it expands in a tight balloon or
loop-like structure.  As the flow evolves each loop interacts with
its neighbor leading to a saturation of the growth.  Consideration
of the poloidal and toroidal fields shows that initially the loops
are dominated by the toroidal components.

%%%%%%%%%%%%%%%%%%
\placefigure{fig:simB}
%%%%%%%%%%%%%%%%%%
%%%%%%%%%%%%%%%%%%
\placefigure{fig:simDRAWING}
%%%%%%%%%%%%%%%%%%
In Fig.~\ref{fig:simDRAWING}, the gray-scale map of the density, we have drawn
schematically the envelope of the largest instabilities (dashed line)
and marked by large arrows the corresponding location of the knots.
The smallest internal instabilities are also shown by smaller grey arrows.
We note that there exist globally five small instabilities between two
consecutive knots of the envelope. The corresponding wavelengths are
reported in table~\ref{table:sim}. The corresponding wavenumbers are also provided.

%%%%%%%%%%%%%%%%%%
\placetable{table:sim}
%%%%%%%%%%%%%%%%%%
We note that the wavelength of the envelope is around $3 r_{jet}$ and that the
smallest instabilities correspond to half the jet radius.
As shown by Appl \etal (\cite{lba}), current driven instabilities
are absolute instabilities, \ie they grow but do not propagate,
in the rest frame of the jet, and the boundary conditions at the
outer edge of the jet do not affect the internal modes.
Hence internal instabilities will develop first as seen
in Fig.~\ref{fig:simDRAWING}. On the other hand, the instabilities with
large wavelengths that develop on the envelope will directly depend on the
the boundary of the jet that is only perturbed when the internal instabilities
become important. Thus {\em in real astrophysical jets, small internal
instabilities should develop before larger ones which give rise
to the larger pinches of the jet}.

%%%%%%%%%%%%%%%%%%%%%%%%%%%%%%%%%%%%%%%%%%%%%%%%%%%%%%%%%%
\subsection{Comparison of the Results}
%%%%%%%%%%%%%%%%%%%%%%%%%%%%%%%%%%%%%%%%%%%%%%%%%%%%%%%%%%

Since our simulation is axisymmetric and therefore only track the pinch
modes, the results of the simulation can only be compared with those
presented in section \ref{sec:m=0}.

In the axial part of the jet, the rotation parameter $\omega$ is
small with respect to unity. Following the classification of paper I,
this corresponds to the characteristic of a slow rotator.  Hence, for
our simulation, the instabilities in the core should be related to the
results presented in table~\ref{table:m0} for the slow rotator. On the other
hand, in the intermediate and the outermost regions, the jet properties
correspond to those of a fast rotator. The largest growth rate
for the fast rotator corresponds to that of the envelope.
%%%%%%%%%%%%%%%%%%
\placetable{table:ratio}
%%%%%%%%%%%%%%%%%%
These wavelengths have been reported in table~\ref{table:ratio},
together with their ratio, and show good agreement.

The first internal instabilities arise at around $5~r_{jet}$ in
the simulation. They grow and become important at about
$10~r_{jet}$. The mean time for the instabilities to grow is then
approximatively given by
\begin{equation}
\tau_I\approx 5 \ r_{jet}/v_{jet} .
\end{equation}
Let us compare
this with the growth time of the pinch mode instabilities $\tau_{slow}$
for the slow rigid rotator. The inverse of the maximum growth rate for
 slow rotator (table~\ref{table:m0}) gives it as
$\tau_{slow}=\Gamma_{slow}^{-1}=1.5 \ \tau_A$, where $\tau_A=r_{jet}/v_A$
is the Alfv\'en time. Using the value of the relative velocity in
Fig.~\ref{fig:ASYMP} (upper left panel), we can derive the velocity of
the jet w.r.t. the Alfv\'en velocity, \ie $v=v_{jet}/ v_{A}\approx 3\pm0.5$.
Then the Alfv\'en time is approximately given by
$\tau_A\approx 3(\pm0.5) r_{jet}/v_{jet}$. Hence we finally have
\begin{equation}
\tau_{slow}=1.5\ \tau_A\approx 4.5 \ (\pm0.75) \ r_{jet}/v_{jet},
\end{equation}
that can be compared to $\tau_I$. Hence the instabilities develop in
the simulation as predicted by the stability analysis. Moreover these
growth times are rather short and consequently the internal instabilities
in astrophysical jets should develop relatively close to the source
(at about $5~r_{jet}$).
\footnote{
As shown by Begelman (\cite{begelman}), the distance $R_{min}$
from the source where instabilities start growing in a non-relativistic,
supersonic jet must be larger than $M_A~r_{jet}$ (in our case
$R_{min}>2.5~r_{jet}$).
}

Thus the theoretical results are in good agreement with the simulations.
The present stability analysis seems to be a good diagnostic
tool to understand the development of instabilities in MHD jets, and
to predict their characteristic wavelengths.

%We note, however, that
%the real nature of the instabilities, \ie KH, CD or PD, has not been
%clarified, even if the CD instabilities might be the most important
%due to the strong toroidal magnetic field.

%%%%%%%%%%%%%%%%%%%%%%%%%%%%%%%%%%%%%%%%%%%%%%%%%%%%%%%%%%
\section{Conclusions}
%%%%%%%%%%%%%%%%%%%%%%%%%%%%%%%%%%%%%%%%%%%%%%%%%%%%%%%%%%

In this work we have investigated Keplerian MHD jet equilibria
by means of a simplified axisymmetric, polytropic and stationary model.
It assumes that the magnetic surfaces possess a shape, conical
in the present case, which is known a priori inside the fast
critical surface. In this {\em inner region},
the problem is entirely specified by $\Omega(a)$, $Q(a)$, and $\alpha_0$.
The Alfv\'en regularity condition together with the criticality
conditions determine the three other unknown first integrals: $E(a)$, $L(a)$
and $\alpha(a)$. In the {\em asymptotic region}, the flow is uniquely
determined by the first-integrals obtained in the {\em inner region}
together with the axial density $\rho_0$ which is given as boundary
condition. This region is governed by the asymptotic forms of the
Bernoulli and transfield equations. Our principal conclusions can be
summarized as follows:
\begin{enumerate}
\item In the {\em inner region}, we find that the fast magneto-sonic
and Alfv\`en surfaces strongly inflate when rotation increases, while
the slow surface gets smaller. The fast point remains at a finite
distance for finite entropy flows, in contrast to cold flows.
\item In the {\em asymptotic region}, jets have a dense and
current-carrying central core where most of the momentum is located.
Rigid body rotation jets are  surrounded by a tenuous current-free
envelope, dominated by the azimuthal magnetic field. {\em Keplerian} and
{\em multi-fast} jets are surrounded by a denser collar and carry an
internal return current.  The intermediate part of the jet is almost
current-free and is magnetically dominated.
\item An approximate analytical solution was derived which may make it possible to
estimate the properties of the outflow and of the emitting object
directly from the quantities observed in jets, {\em e.g.} the
density, the components of the jet velocity.
\item We have addressed the linear stability and the non-linear
development of instabilities for the corresponding equilibria
using both analytical and 2.5-D numerical simulations.
Instabilities in the simulations develop with a wavelength and growth
time that are well matched by the stability analysis.
Regardless the type of rotators, the wavelengths of the axisymmetric
modes suggest that these instabilities could be at the origin of the
knotted morphology of a large number of astrophysical jets.
\item The rather short growth times of the internal instabilities
suggest that they should develop relatively close to the source,
at about $5~r_{jet}$ for YSO jets.
\item We have shown that ambipolar diffusion should be effective
on times scales of order $10^4$ to $10^5$ y, or a length scale equal
or larger than $0.1 ~pc$, at least in some part of typical YSO jets.
The consequent changes of the internal magnetic configurations
might be the reason for the disappearance of knots
at such a distance from the source, as observed.
\item
The relatively good agreement between the theoretical results and the
simulation shows that the present stability analysis is a good diagnostic
tool to understand the development of instabilities in MHD jets.
\end{enumerate}

%-------------------------------------------------------------------------
\acknowledgements

We wish to thank Hubert Baty, Stefan Appl, Jean Heyvaerts
for their help with this project.  We thank Tom Gardiner and Guy
Delamarter for useful discussions and Tom Jones and Dongsu Ryu
for their help with the numerical code.  This work was supported by NSF Grant
AST-0978765 and by the University of Rochester's Laboratory for
Laser Energetics. The project was also supported by an operating 
grant from NSERC of Canada.

%-------------------------------------------------------------------------

\appendix
\section{The Differential Equations of the Model}

%%%%%%%%%%%%%%%%%%%%%%%%%%%%%%%%%%%%%%%%%%%%%%%%
1. The so-called {\em Alfv\'en regularity condition} is the particular 
form assumed by the force balance perpendicular to the magnetic
field, or transfield  equation, at the Alfv\'en point.
It can be written as
\begin{eqnarray}
{{\alpha'}\over{\alpha}}  + 2 (1-p) {{r'_A}\over{r_A}}
- 2 (1-p) {{sin \theta_A}\over{r_A \vert \nabla a \vert_A}}
-  {{Q \rho_A^{\gamma -1}}\over{(\gamma - 1)
v_{PA}^{2}}}{{Q'}\over{Q}}&&
\nonumber \\
+{{E'}\over{v_{PA}^{2}}}
+ {{\Omega^2 r_A^2}\over{v_{PA}^{2}}}
\left( {{\alpha'}\over{\alpha}} {{1}\over{(1-p)^2}}
+ 2 {{r'_A}\over{r_A}} {{p}\over{1-p}}
 -{{\Omega'}\over{\Omega}}\right) &= &0 ,
\label{RegAL}
\end{eqnarray}
where $p$ is the slope of the solution of the Bernoulli equation at the
Alfv\`en point.  The reader should see \cite{HeyNor89},
\cite{lery1} and \cite{lery2} for a more detailed explanation of the
the equations and methods

%%%%%%%%%%%%%%%%%%%%%%%%%%%%%%%%%%%%%%%%%%%%%%%%
2. The projection of the equation of motion on ${\bf B}_P$
yields by integration the {\em Bernoulli equation},
$E(a)={{v^2}/{2}} + \left({{ \gamma}/(\gamma - 1)}\right) 
Q \rho^{\gamma - 1} + G(r,z) - {{r \Omega B_{\theta}}/{\mu_0 \alpha}}$. 
In the absence of MHD forces, this first integral would express 
the well known Bernoulli's theorem, i.e. the constancy of 
the sum of the kinetic, enthalpy and gravitational energy 
fluxes. The presence of the magnetic field introduces another
energy flux, the Poynting flux, the fourth term in the equation. 
The total specific energy has a given value on a  magnetic surface.
This can be expressed in differential form as
\begin{equation}
{d \over da} (E_s - E_f) = 0 .
\label{NUM3}
\end{equation}
where the $s$ and $f$ subscripts refer to the slow and fast
points respectively

%%%%%%%%%%%%%%%%%%%%%%%%%%%%%%%%%%%%%%%%%%%%%
3. Every solution has to fulfill {\em four criticality conditions}
defined at the slow and fast magneto-sonic points. These points are 
located where the differential with respect to $\rho$ and $r$
of the Bernoulli function ${\mathcal B}(r,\rho)=E(a)$ vanishes, \ie
$r {\partial  {\mathcal B}}/{\partial r }= 0$,
$\rho {\partial {\mathcal B} }/{\partial \rho} = 0$.
It is convenient to convert the algebraic magnetosonic criticality
conditions into differential equations. The critical points can be 
followed from one magnetic surface to the next by differentiating 
these criticality equations with respect to $a$:
\begin{equation}
\frac{d}{da} \biggl(\frac{\partial {\mathcal B}}
{\partial r}\biggr)_{s} = 0
\qquad,\qquad
\frac{d}{da} \biggl(\frac{\partial {\mathcal B}}
{\partial \rho}\biggr)_{s} = 0
\label{NUM1S}
\end{equation}
\begin{equation}
\frac{d}{da} \biggl(\frac{ \partial{\mathcal B}}
{\partial r}\biggr)_{f} = 0
\qquad,\qquad
\frac{d}{da} \biggl(\frac{\partial {\mathcal B}}
{\partial \rho}\biggr)_{f} = 0 .
\label{NUM1F}
\end{equation}

4. The {\em asymptotic form of the Bernoulli equation} can be written as: 
\begin{equation}
\frac{d r}{d a}  ={ \alpha}
\left[{\rho r \sqrt{2}\sqrt{E-\frac{\gamma}{\gamma-1}
Q \rho^{\gamma-1}-\frac{\Omega^2 r^2 \rho}{\mu_0 \alpha^2}}}\right]
^{-1}.
\end{equation}

%%%%%%%%%%%%%%%%%%%%%%%%%%%%%%%%%%%%%%%%%%%%%
5. The asymptotic form of the transfield equation
is given by:
\begin{equation}
r^2 {d \over da} \left(Q \rho^{\gamma}\right)
+ {1\over {2}}
{d\over da } 
\left({\Omega^2 r^4 \rho^2} \over {\mu_0 \alpha^2}\right)= 0.
\end{equation}

\section{The Stability Analysis Method}

A global normal mode stability analysis is used. 
The radial displacements, $\xi_r$, of the
fluid elements from their equilibrium positions are
of the form, 
\begin{equation}
\xi_r (\vec{x}) = \xi_r(r)\, \exp i(m\phi + kz - \omega t)
\end{equation}
and similarly for the remaining perturbed quantities.
The linearized equations can be cast into a system of two first order ODE
for $Y\equiv r \xi_r$ and the total perturbed pressure 
$Z \equiv \delta p_{\rm tot} = \delta (p + B^2/8\pi)$. It is given by
\begin{equation}
AS {d Y \over d r} = C_1 Y - r C_2 Z,
\end{equation}
and
\begin{equation}
AS {d Z \over d r} = C_3 Y/r  - C_1 Z,
\end{equation}
where $A, S$ and $C_i$ functions of the equilibrium quantities as well
as the
Fourier parameters $\omega, k, m$ (Appl \& Camenzind \cite{applcam}). 

Jets propagate through a compressible medium of high conductivity.
Since one is interested only in instabilities due to the jet itself,
radially outgoing decaying waves for large radii
define the boundary condition,
\begin{equation}
P \propto H_m^{(1)} \left( \lambda r \right)
\end{equation}
for $r \to\infty $. 
The $H_m^{(1)}$ are the Hankel functions of the first kind and
$\lambda^2 = \omega^2/c_{\rm s}^2 - k^2$, with $c_{\rm s}$ the sound speed,
if the ambient medium is assumed to be unmagnetized.
The perturbed jet is assumed to remain in equilibrium
with its surroundings.
Regularity on the axis provides the other boundary condition, \ie
$Y = 0$ for  $r = 0$.

%%%%%%%%%%%%%%%%%%%%%%%%%%%%%%%%%%%%%%%%%%%%%%%%%%%%%%%%%%%%%%%%
\clearpage
\plotone{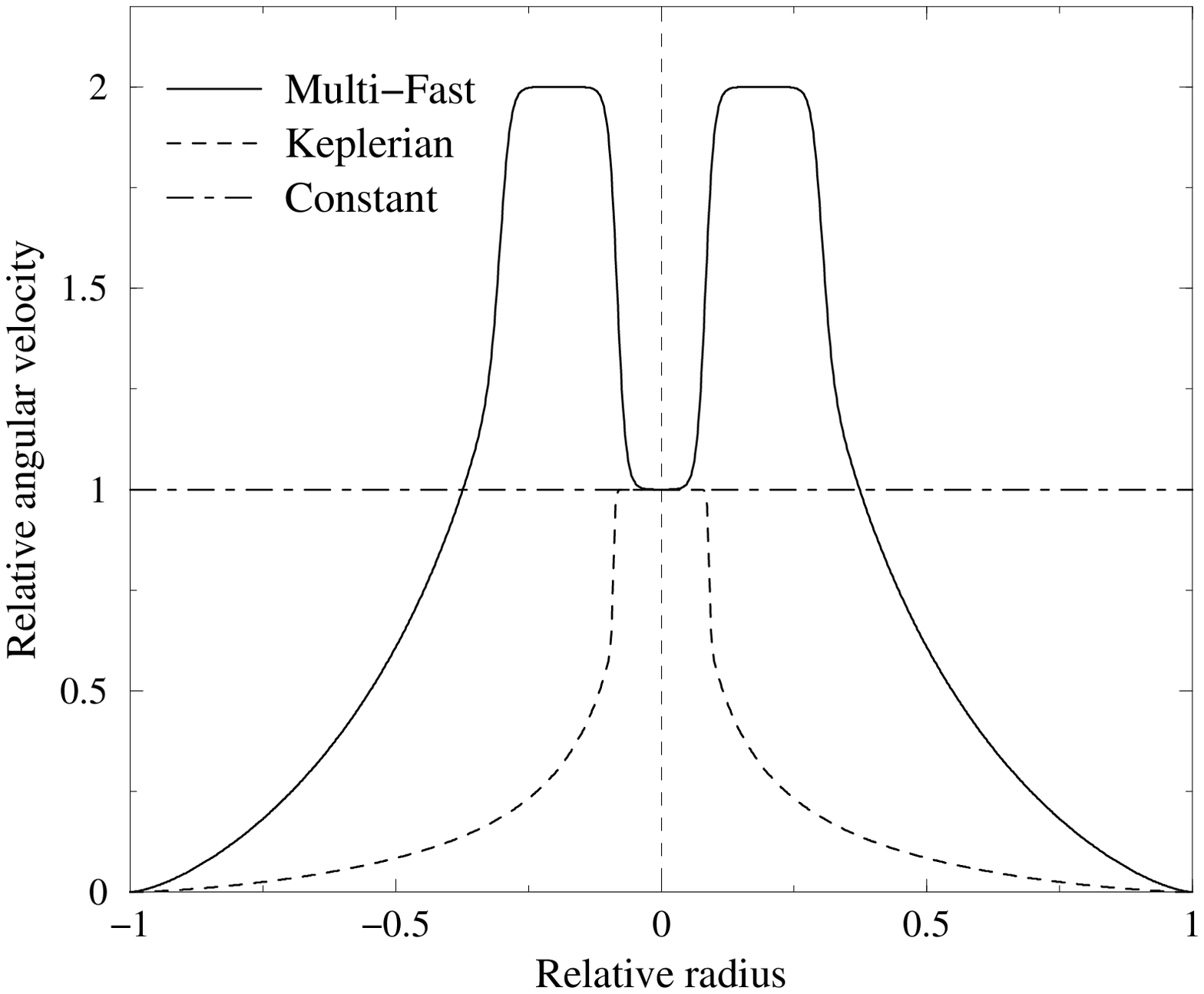}
\figcaption[]{Rotation laws for constant (dot-dashed), pure {\em Keplerian}
(dashed), and multi-components, or {\em multi-fast}, (solid) models.
Axial angular velocity is set to unity.
\label{fig:OMEGA}
}
%\clearpage
\plotone{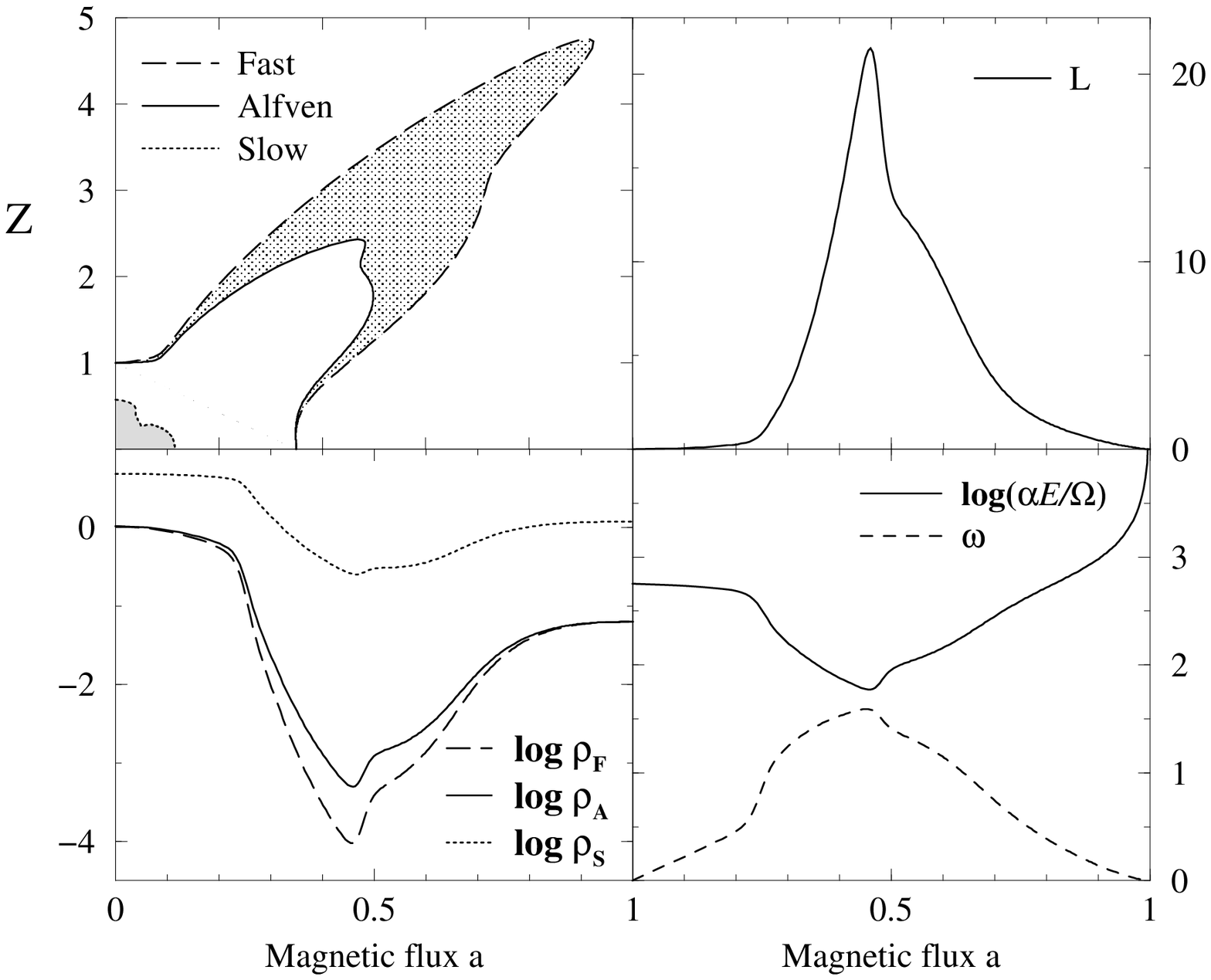}
\figcaption[]{{\em Multi-fast case} :
 We show the critical surfaces in dimension-less quantities (upper left),
and the densities on these surfaces (lower left). Three first integrals,
namely the total angular
momentum $L$ (upper right), the rotation parameter $\omega$ and the
ratio $\alpha E/\Omega$  (lower right) are calculated on the Alfv\'enic
surface as functions of the magnetic flux $a$. Input parameters are
$\overline{Q}=0.87$, $\overline{\Omega}=2$ and $\overline{\alpha_0}=0.7$
($\gamma \approx 1$).
\label{fig:CRIT}}
\clearpage
\plotone{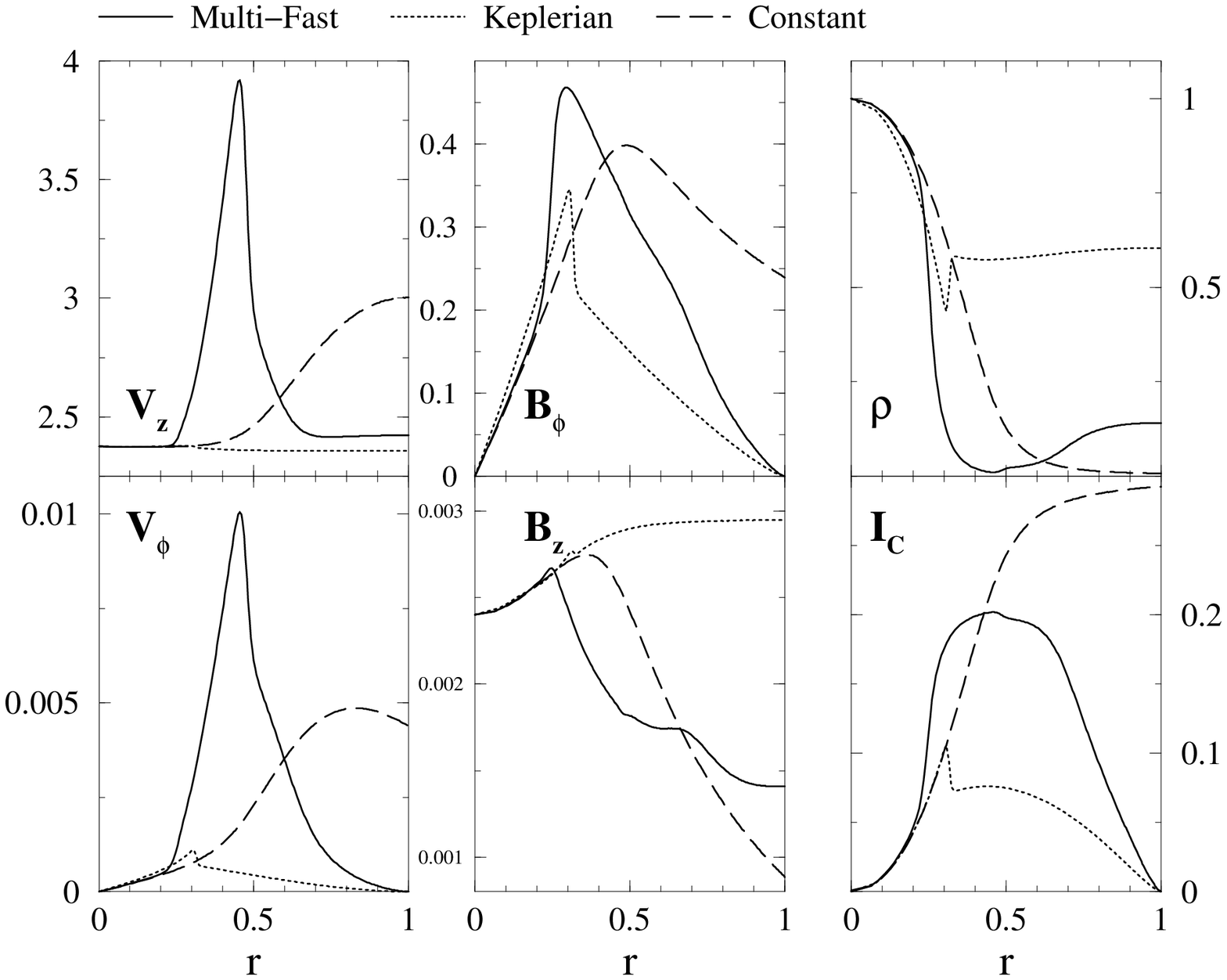}
\figcaption[]{Variations with the relative radius of the velocity $V$
and magnetic field $B$ components, of the density $\rho$ and of the
net electric current $I_C$ in the {\em cylindrically collimated regime}.
Constant (dotted lines), pure {\em Keplerian} (dashed), and
{\em multi-fast} (solid)
rotation laws are considered.
\label{fig:ASYMP}}
\clearpage
\plotone{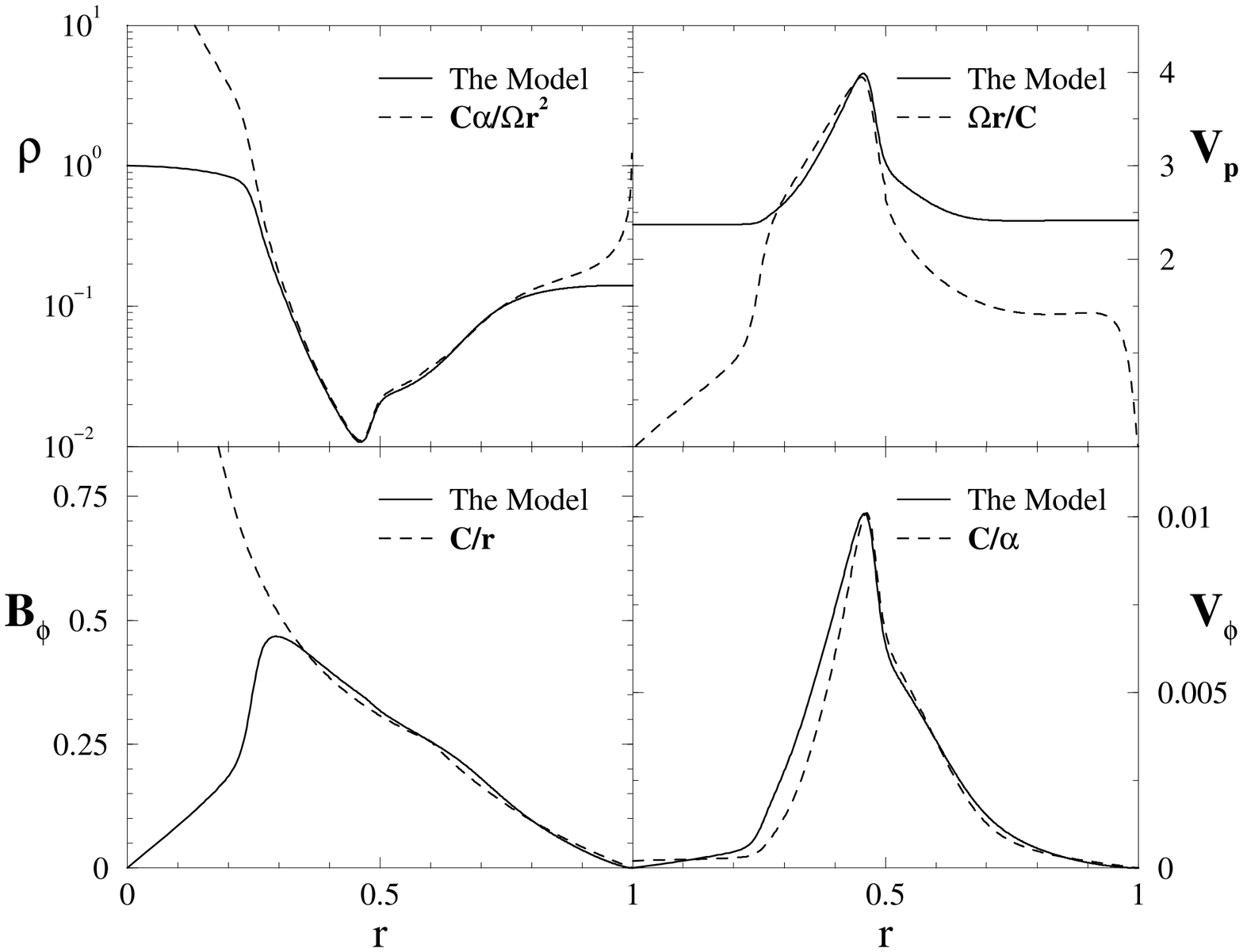}
\figcaption[]{Jet variable distributions for {\em multi-fast} (solid line) and approximate
analytical (dashed line) solutions in the asymptotic region as
functions of the relative radius.
\label{fig:THEO}}
\clearpage
\epsscale{.7}
\plotone{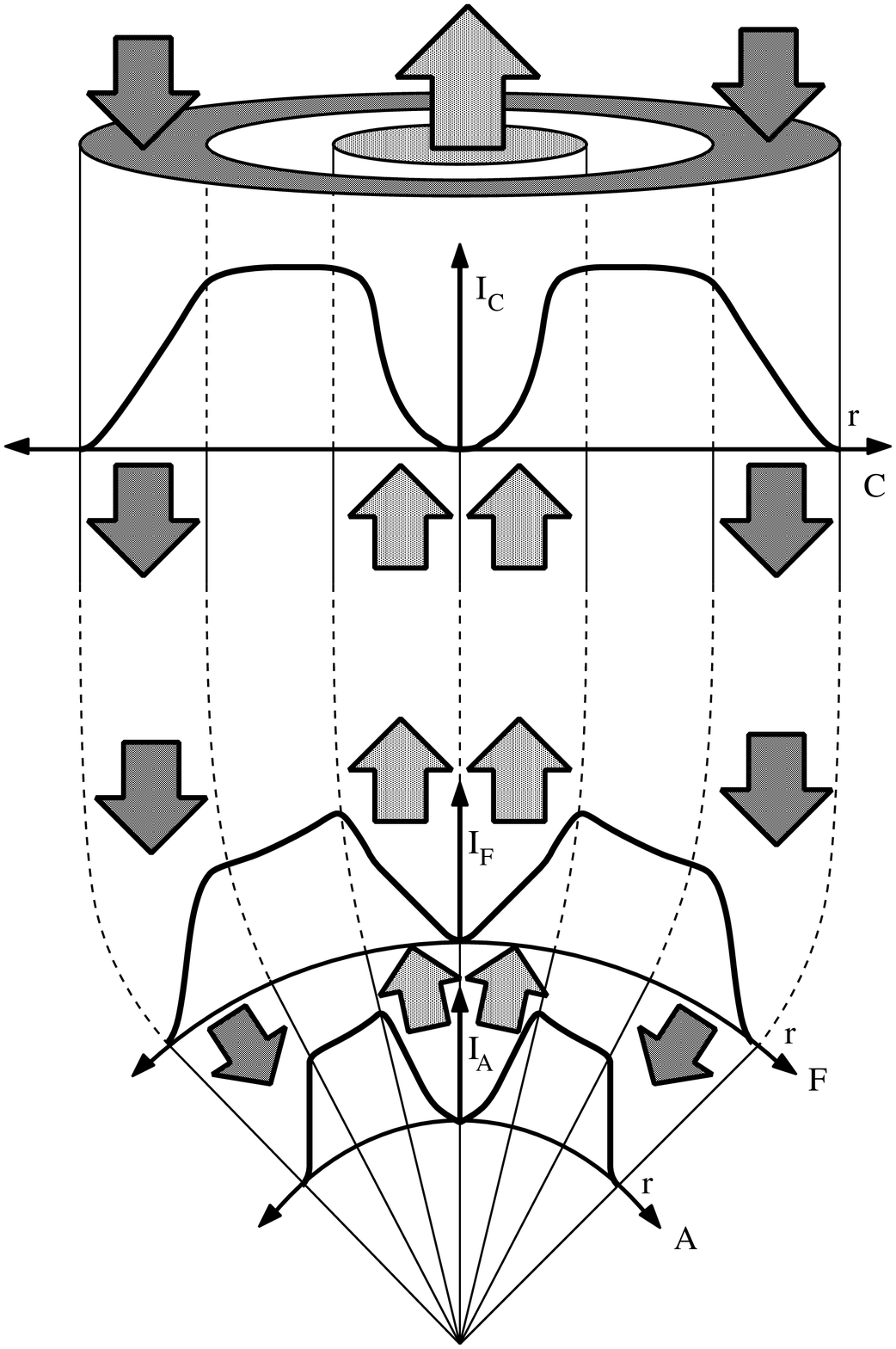}
\figcaption[]{Schematic representation of the variations of the net electric
current along the jet (heavy solid lines), and its direction (arrows), at
the Alfv\'enic (A), and the fast (F) surfaces, and in the cylindrically
collimated regime (C). The light solid (and dashed) lines
represent the poloidal projection of magnetic field lines.
\label{fig:CURRENT}}
\epsscale{1.}
\clearpage
\plotone{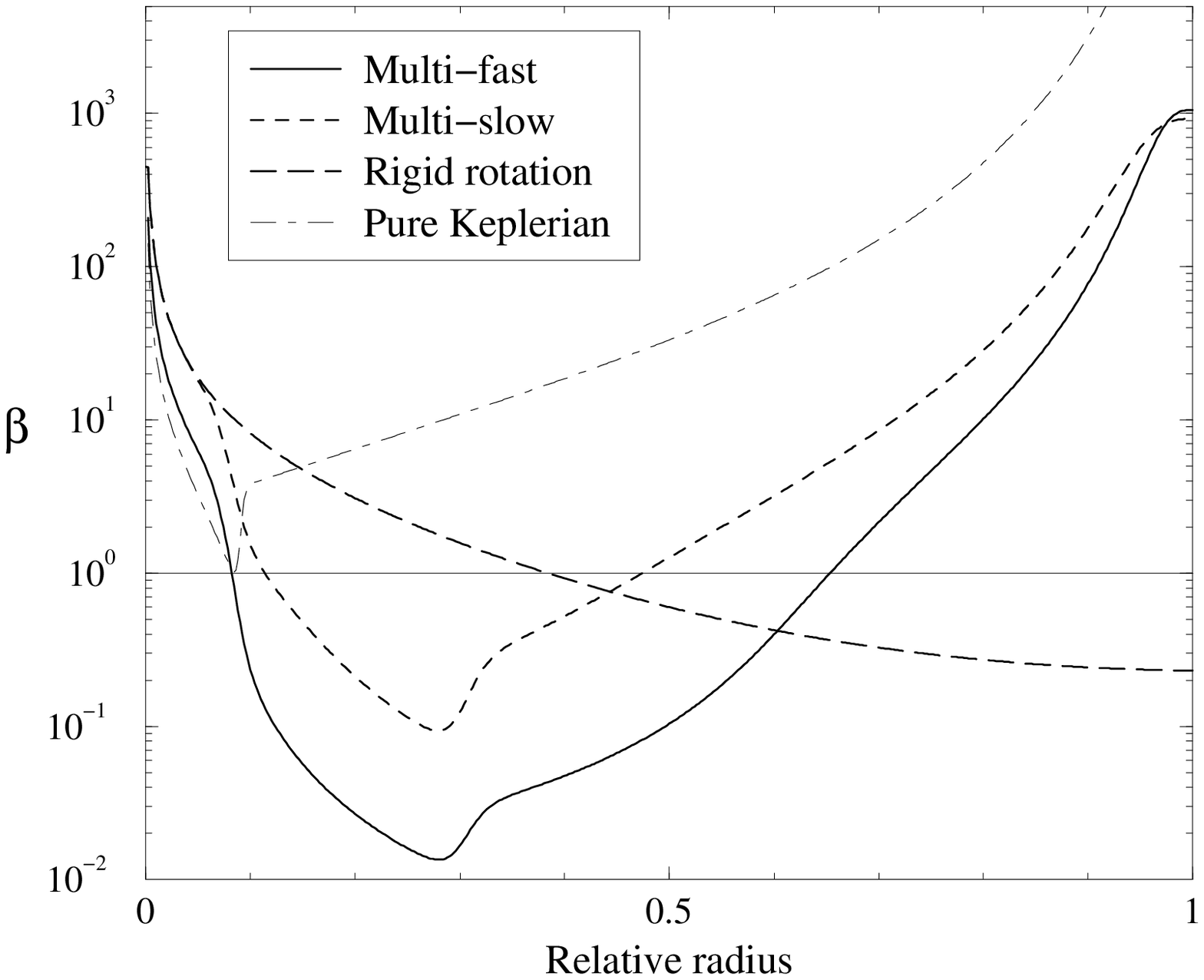}
\figcaption[]{ Plasma $\beta$ parameter as a function of the relative radius
for the {\em multi-fast} (solid), {\em multi-slow} (dashed), rigid body
(long dashed) and pure {\em Keplerian} (dotted-dashed) rotation cases.
\label{fig:BETA}}
\clearpage
\plotone{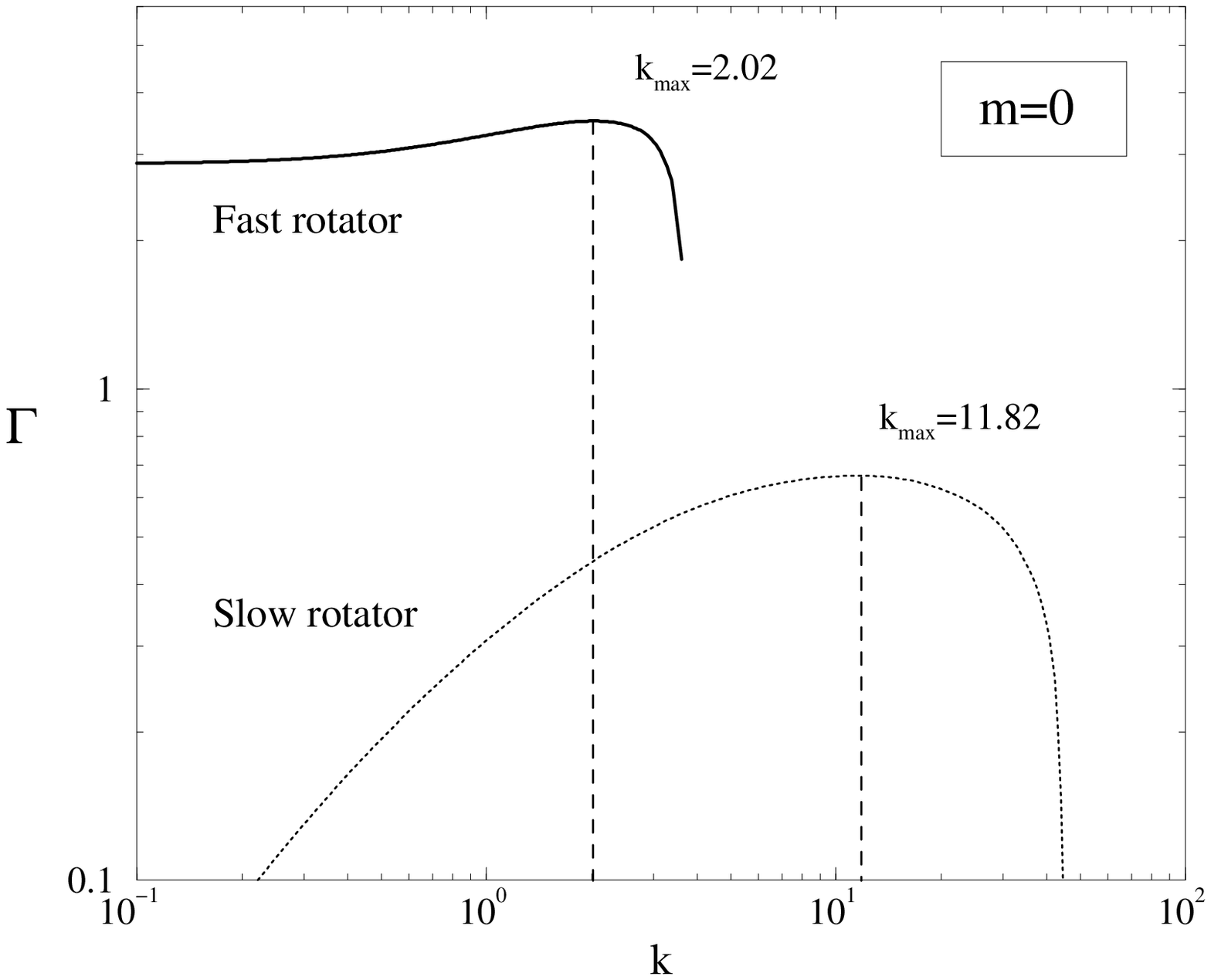}
\figcaption[]{Dispersion relation of the pinch mode ($m=0$)
for fast and slow rotators (solid and dotted lines respectively).
 Dashed lines correspond to the maximum wavenumbers
($k_{max}=2.02$, $k_{max}=11.82$).
\label{fig:m0}}
\clearpage
\plotone{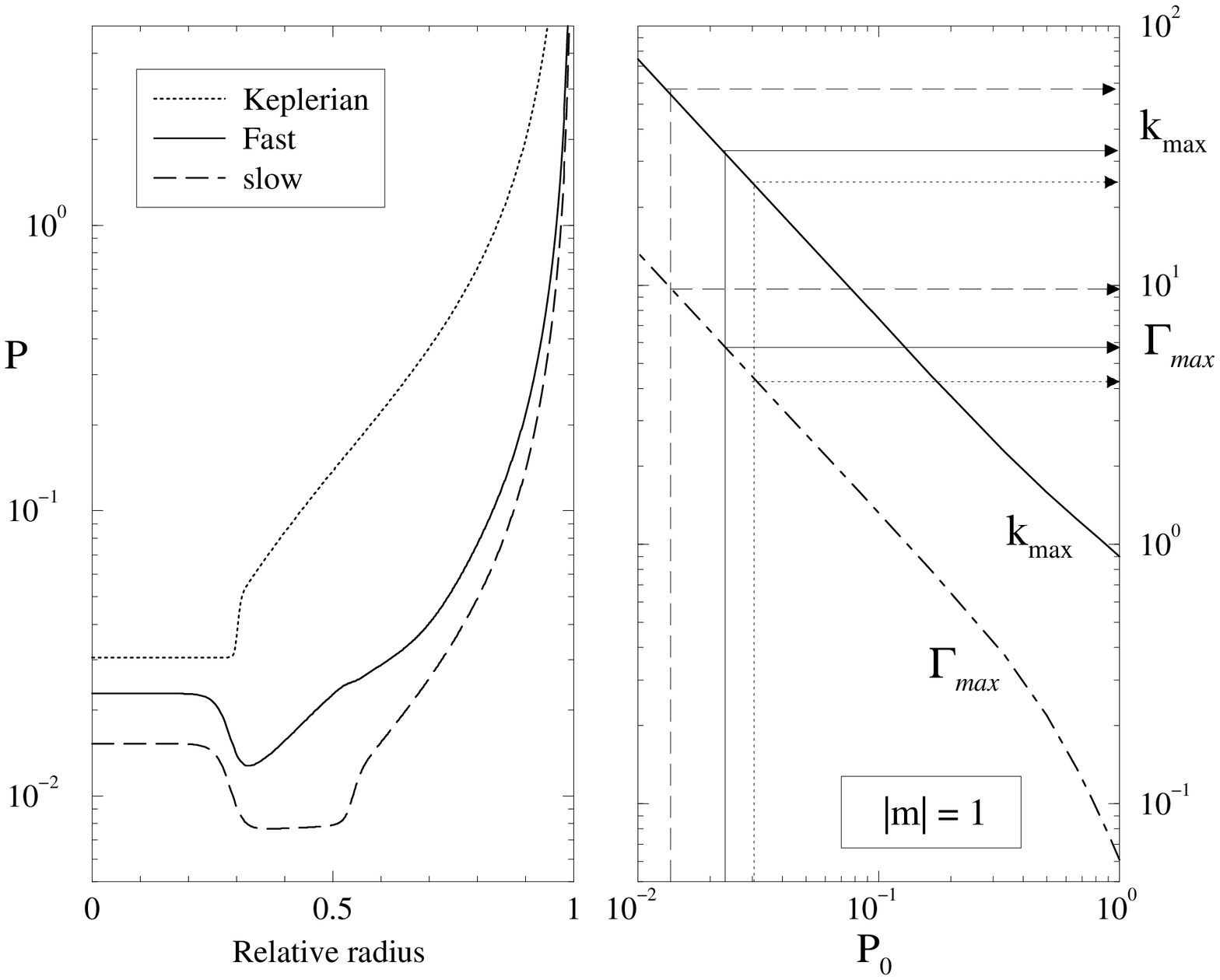}
\figcaption[]{The Pitch profile (left panel) is shown for the {\em Keplerian}
(dotted lines), the {\em multi-fast} (solid), and the {\em multi-slow}
(dashed) cases.
On the right panel, the maximum growth rate ($\Gamma_{max}$) and the
maximum wavenumber ($k_{max}$) are plotted as functions of the central
pitch value. The arrows mark their estimate for the different cases.
 \label{fig:muGK}}
\clearpage
\plotone{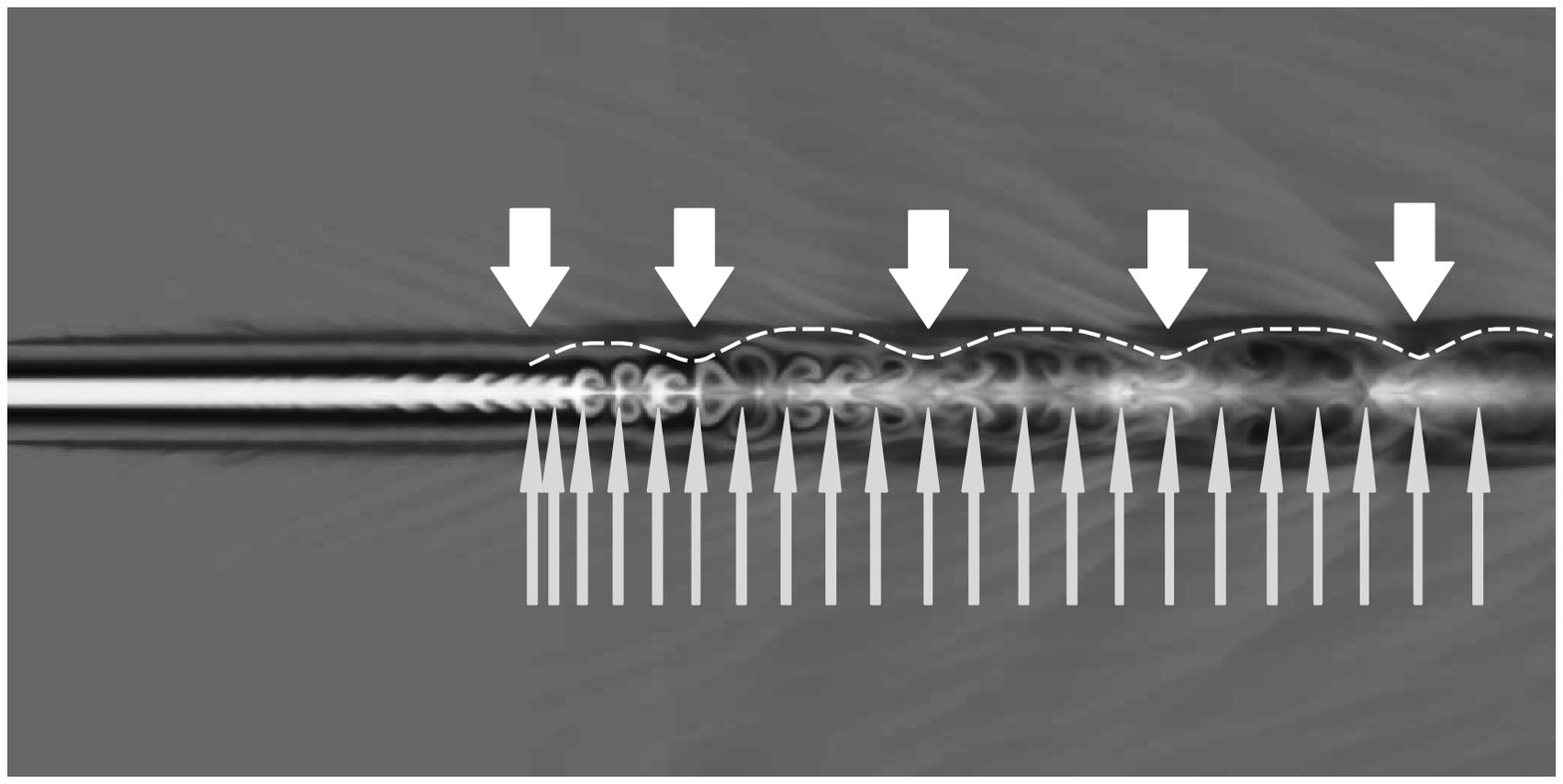}
\figcaption[]{Structures of instabilities.
Thick white arrows correspond to knots of the envelope (delimited by
dashed line), and thin black arrows to maxima of density for the internal
instabilities.
\label{fig:simDRAWING}
}
\plotone{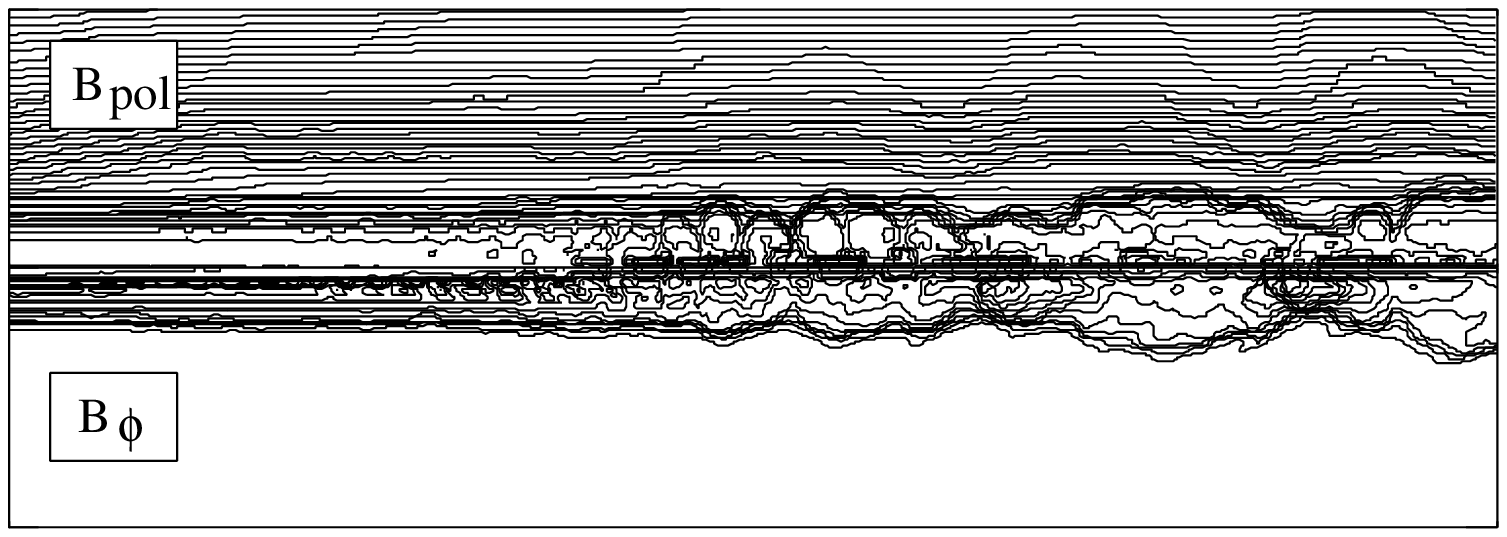}
\figcaption[]{
Components of the magnetic field of the destabilized equilibrium
\label{fig:simB}
}
%%%%%%%%%%%%%%%%%%%%%%%%%%%%%%%%%%%%%%%%%%%%%%%%%%%%%%%%%%%%%%%

\clearpage
%%%%%%%%%%%%%%%%%%%
\begin{deluxetable}{cccc}
\footnotesize
\tablecaption{The Pinch Mode:
growth rate $\Gamma_{max}$, wave number $k_{max}$, and wavelength
$\lambda_{max}$  of the instability with the largest growth rate for
the slow and the fast rigid body rotators.
\label{table:m0}
}
\tablewidth{0pt}
\tablehead{
\colhead{type}
& \colhead{$\Gamma_{max}$}
& \colhead{$k_{max}$}
& \colhead{$\lambda_{max}$}
}
\startdata
Fast rotator & 3.50 & 2.02  & 3.11  	\nl
Slow rotator & 0.67 & 11.82 & 0.53  	\nl
\enddata
\end{deluxetable}

%%%%%%%%%%%%%%%%%%%
\begin{deluxetable}{ccccc}
\footnotesize
\tablecaption{
The Helical Mode:
central pitch $P_0$, growth rate $\Gamma_{max}$, wavenumber $k_{max}$,
and wavelength $\lambda_{max}$ of the instability with the largest growth
rate for the different rotators.
\label{table:m-1}
}
\tablewidth{0pt}
\tablehead{
\colhead{type}
& \colhead{$P_0$}
& \colhead{$\Gamma_{max}$}
& \colhead{$k_{max}$}
& \colhead{$\lambda_{max}$}
}
\startdata
{\em Keplerian}	& 0.031	& 4.2 & 25 & 0.25 \nl
{\em multi-fast}	& 0.023	& 5.5 & 30 & 0.20 \nl
{\em multi-slow}	& 0.015	& 9.5 & 55 & 0.11 \nl
\enddata
\end{deluxetable}

%\clearpage
%%%%%%%%%%%%%%%%%%%
\begin{deluxetable}{crrrr}
\footnotesize
\tablecaption{
The simulation: number of instabilities $N$, their size (in units
of $r_{jet}$), their mean wave number $\bar k=2\pi N/size$, and
the mean wavelength $\bar \lambda=size/N$ for the
{\em multi-fast} case.
\label{table:sim}
}
\tablewidth{0pt}
\tablehead{
\colhead{type}
& \colhead{$N$}
& \colhead{size}
& \colhead{$\bar k$}
& \colhead{$\bar \lambda $}
}
\startdata
Envelope & 3  & 8.6  $\pm$ 0.75 & 2.19 $\pm$ 0.19 & 2.87 $\pm$ 0.25 \nl
Core	 & 21 & 10.5 $\pm$ 0.75 & 12.57 $\pm$ 1.00 & 0.50  $\pm$ 0.04\nl
\enddata
\end{deluxetable}

%%%%%%%%%%%%%%%%%%%
\begin{deluxetable}{clll}
\footnotesize
\tablecaption{
Wavelengths of the envelope $\lambda_E$, of the core $\lambda_C$,
and their ratio.
\label{table:ratio}
}
\tablewidth{0pt}
\tablehead{
\colhead{method}
& \colhead{$\lambda_E$}
& \colhead{$\lambda_C$ }
& \colhead{ratio }
}
\startdata
Stability analysis & 3.11 	& 0.53  	&  5.87	\nl
Simulation	& 2.87$\pm$0.25 & 0.50$\pm$0.04 &  5.74$\pm$0.96\nl
\enddata
\end{deluxetable}
%%%%%%%%%%%%%%%%%%%

\end{document}